\documentclass{aastex631}
\usepackage{microtype}
\usepackage{multirow}
\usepackage{gensymb}
\usepackage{xcolor}
\usepackage{amsmath}
\usepackage[bottom]{footmisc}
\usepackage{comment}
\usepackage{float}
\usepackage{xcolor}

\newcommand{\tr}{\textcolor{black}}
\newcommand{\tb}{\textcolor{black}}
\begin{document}
\title{Evolution and mergers of eccentric white dwarf binaries encountering intermediate-mass black holes}
\correspondingauthor{Tapobrata Sarkar}
\email{tapo@iitk.ac.in}

\author{Adarsh Pandey}
\affiliation{Department of Physics, Indian Institute of Technology Kanpur,
Kanpur 208016, India}

\author{Aryabrat Mahapatra}
\affiliation{Department of Physics, Indian Institute of Technology Kanpur,
Kanpur 208016, India}

\author{Tapobrata Sarkar}
\affiliation{Department of Physics, Indian Institute of Technology Kanpur,
Kanpur 208016, India}

\begin{abstract}
	We study how a close passage past an intermediate-mass black hole (IMBH) changes the evolution of an eccentric white dwarf (WD)
	binary that has nominal Roche lobe overflow at internal pericenter. Using three-dimensional smoothed particle hydrodynamics, we follow binaries with
	eccentricity $e_{\rm in}=0.6$ and mass ratios $q=0.2$ -- $0.7$ on parabolic, retrograde orbits around a Schwarzschild IMBH at two
	encounter strengths, $\beta^b=0.5$ and $1.5$. We compare them with the same binaries evolved in isolation. The IMBH greatly
	speeds up the merger. Every IMBH-perturbed binary that merges does so during or shortly after its second internal pericenter passage,
	which is always less than the time taken to merge in isolation. 
	The merger outcome is not monotonic with encounter strength. The $q=0.2$ binary retains two bound cores for $\beta^b=1.5$ but merges at
	$\beta^b=0.5$. In the deeper encounter it settles onto a less eccentric post-encounter orbit, showing that the timing of the
	encounter relative to the inner orbit plays an important role. Deep encounters also produce transient reversals of the
	internal orbital angular momentum for $q=0.2$ -- $0.4$, while the material bound to the accretor gains more spin angular
	momentum than that bound to the donor. The specific orbital angular momentum of the binary CM around the IMBH changes by only a
	few $\times10^{-4}$ of its initial value. We also identify He-origin material that can burn helium on a dynamical timescale,
	amounting to $\sim15$ -- $17\%$ of the He-origin mass for $q=0.7$. IMBH encounters can therefore quantitatively change the
	evolution of eccentric WD binaries.
\end{abstract}

\section{Introduction}
\label{sec1}

Compact white dwarf (WD) binaries are among the most common compact object systems in the Galaxy. They are main sources
for low-frequency gravitational-wave (GW) detectors \citep{Amaro-Seoane2017} and possible progenitors for several kinds of thermonuclear
transients \citep{Maoz2014}. Population synthesis predicts $O(10^{8})$ such systems in the Milky Way
\citep{Nelemans2001}, and binaries with orbital period 7 minutes have been observed \citep{Burdge2019}.

In dense stellar environments such as globular clusters and galactic nuclei, stellar encounters, mass segregation, and
hierarchical perturbations can modify binary orbits and produce large eccentricities
\citep{Ivanova2006,Ivanova2010,Samsing2014,Naoz2016}. Repeated encounters with passing stars can further harden
already compact binaries and modify their orbital dynamics \citep{Heggie1975,Hut1983}. Mass segregation can also bring relatively
massive binaries into the central regions of stellar systems \citep{Freitag2006,AlexanderHopman2009}, where a central massive
black hole (BH) can further alter the inner orbit of a bound binary through secular perturbations, including Kozai--Lidov eccentricity oscillations
\citep{Antonini2012,Naoz2016,Stephan2016}. Further, studies of WD tidal capture by intermediate-mass black holes (IMBHs, $10^2 < M_{\rm BH}/M_\odot < 10^6$) show
that WDs can be supplied to the immediate environments of IMBHs in dense stellar systems \citep{Ye2023}. Hence, eccentric compact
binaries interacting with central IMBHs are a natural possibility in dense stellar environments.

A binary responds to a BH tidal field differently from a single star. Because the internal binary orbit is weakly bound
compared to stellar interiors, an encounter can change the binary separation and eccentricity, while leaving the
individual stars intact.
Depending on the encounter geometry, the binary pair may be separated or widened, or driven into closer interaction. 
In this paper we consider an eccentric binary that is undergoing or close to Roche lobe overflow (RLOF) when the tidal field of the BH acts on it. 
We study the mass transfer and possible mergers that follow. Hydrodynamic studies of binaries near massive BHs have so far 
treated detached systems, and the effect of an external tidal field on a binary that is already transferring mass has not
been studied. This is what our paper aims to do. 

Hydrodynamic encounters between individual WDs and massive BHs have
been studied extensively. For non-spinning BHs, tidal disruption of a
WD outside the event horizon is restricted primarily to the IMBH mass regime.
Numerical simulations have explored the associated tidal compression, heating,
mass loss, and possible thermonuclear ignition
\citep{Rosswog2009,MacLeod2016,Anninos2018}. Hydrodynamic studies of
interacting double WD systems have mostly considered two cases. 
The first is the quasi-circular inspiral, in which GW
emission brings the binary into Roche lobe contact
\citep[e.g.,][]{Benz1990,RasioShapiro1995,Dan2011,Dan2012,Dan2014,
	Pakmor2010,Pakmor2012,Sato2015}. The second is the dynamically driven,
single-passage collision of two initially unbound or weakly bound WDs, where
strong compression and shock heating can lead to thermonuclear burning
\citep{Rosswog2009Collision,Raskin2009}. A bound eccentric WD binary that transfers
mass near each internal pericenter lies in between these two cases.

If such a binary passes close to a massive BH, the tidal field of the BH acts on 
it while mass transfer is ongoing. The encounter can then change both the inner
orbit and the timing and strength of the later mass transfer episodes. 

Closely related problems have been studied in several ways. Studies treating binary stars as point masses during encounters
with a single massive BH have examined binary breakup, exchange,
capture, and high-velocity ejection
\citep{Hills1988,Sari2010,Kobayashi2012,YuLai2024}, while recent relativistic
extensions indicate that strong-field effects can enhance the probability of
stellar collisions and mergers \citep{Manzaneda2024}. Secular and population
studies of hierarchical WD systems have likewise shown that external
perturbations can excite large eccentricities and promote mergers
\citep{Thompson2011,Hamers2013,Rajamuthukumar2023}, although direct WD--WD
collisions in isolated triples appear to be rare \citep{Toonen2018}.
A smaller number of studies have followed stellar binaries
hydrodynamically during close encounters with massive BHs
\citep[e.g.,][]{Antonini2011,Mainetti2016,YuLai2025}. 
Our problem combines binary -- BH encounter dynamics along with the hydrodynamics of
the interacting binaries, so the tidal field of the BH and the response of the stars
need to be followed together.

In this paper, we perform three dimensional smoothed particle hydrodynamics (SPH) simulations of eccentric double WD binaries on
parabolic encounters with a Schwarzschild IMBH, using the numerical framework of \citet{Banerjee2023}. Stellar hydrodynamics and self-gravity are treated in a Newtonian manner, while the
external orbital motion and relativistic tidal field are evaluated in a fixed Schwarzschild spacetime. In our previous studies
\citep{Binary1,Binary2}, we considered initially detached, circular WD binaries undergoing a single strong encounter. Here we instead
focus on a class of eccentric binaries that undergo RLOF at internal pericenter. We survey four mass ratios and two encounter
strengths and evolve corresponding isolated binaries as a reference. The simulations address three main questions, a) how an IMBH
encounter alters the intrinsic evolution and merger timescale of an eccentric, mass transferring WD binary, b) how angular
momentum is redistributed among the internal orbit, stellar spins, and external center of mass (CM) motion, and c) how the response
depends on binary mass ratio and encounter strength, including the role played by the inner orbital phase at which the binary reaches outer pericenter. 
We find that the IMBH can speed up mass transfer and merger in most configurations. A deeper encounter does not always lead to merger, 
and the outcome depends on the inner orbital phase. Internal orbital angular momentum has a large change, whereas external CM orbital angular momentum changes are small. 
We also study the thermodynamics of the He-origin material and identify configurations where He burning can take place on a dynamical timescale.

The remainder of this paper is organized as follows. Section~\ref{sec:setup} defines the binary--IMBH configuration and model
parameters, and Section~\ref{sec:numerics} describes the numerical method. Section~\ref{sec:results} presents the hydrodynamic
evolution and associated phenomena, while Section~\ref{discussion} discusses the broader implications of the results and
summarizes our conclusions. The thermodynamic burning conditions are presented in Appendix~\ref{nuclear_ignition}. 

\section{Physical Setup}
\label{sec:setup}

We consider a hierarchical system consisting of an inner stellar binary orbiting a central BH. We use subscripts ``in'' and
``out'' to distinguish, respectively, the orbital parameters of the stellar binary from those of the binary CM orbit
around the BH. In our notation, $(a_{\mathrm{in}},\, e_{\mathrm{in}},\, r_{p,\mathrm{in}})$ denote the semi-major axis,
eccentricity, and pericenter distance of the stellar binary, while $(a_{\mathrm{out}},\, e_{\mathrm{out}},\, r_{p,\mathrm{out}})$
denote the corresponding quantities for the orbit of the binary CM around the BH.

\subsection{Binary System}\label{subsec:Binary system}

We consider a binary system composed of two stars with masses $m_1$ and $m_2$, modeled using SPH. The binary is initialized on an
eccentric orbit with the stars placed at the apocenter of their mutual orbit, ensuring a dynamically relaxed initial configuration
free of mass transfer or tidal deformation. The binary pericenter is chosen so that the stars interact strongly at
pericenter passage. As an initial estimate of the Roche lobe scale, we use the Eggleton relation \citep{Eggleton1983},
\begin{equation}
	\frac{r_L}{a_{\rm in}}
	=
	\frac{0.49 q^{2/3}}
	{0.6 q^{2/3}+\ln(1+q^{1/3})},
\end{equation}
where $q=M_2/M_1$ is the donor-to-accretor mass ratio. For an eccentric, nonsynchronous binary, however, the instantaneous
Roche lobe geometry differs from the circular, synchronously rotating case assumed by the Eggleton prescription
\citep{Sepinsky2007a, Sepinsky2007b}, since the effective Roche lobe depends on orbital phase and the degree of
asynchronism. We use the Eggleton relation only to set the scale of the pericenter separation, not as an exact RLOF criterion.
Specifically, the initial pericenter distance is set according to
\begin{equation}
	r_{p,\mathrm{in}}
	=
	\frac{R_2}
	{n\,(r_L/a_{\rm in})},
\end{equation}The parameter $n$ specifies the degree of nominal Roche lobe
overfilling relative to the Eggleton-based reference scale at
pericenter, such that $R_2=n\,r_{L,p}^{\rm Egg}$. Test simulations
with $n=1$ showed that mass transfer in the isolated eccentric binary
is weak and requires many pericenter passages to produce significant
orbital evolution. We therefore adopt $n=1.1$, corresponding to a
10\% nominal overfilling of the Eggleton-based reference scale, in
order to obtain appreciable episodic interaction over a tractable
number of binary orbits. 
Larger values of $n$ lead to deeper initial interactions and these are
not explored here. The parameter $n$ sets the initial orbit, with the onset
and strength of mass transfer then determined by SPH evolution. 

The stellar binary is initialized using the Keplerian two-body solution corresponding to the chosen inner orbital eccentricity
\(e_{\mathrm{in}}\) and pericenter distance \(r_{p,\mathrm{in}}\). For a given true anomaly \(\nu_0\) (defined as the angle in the
orbital plane between the pericenter direction and the separation vector), the instantaneous binary separation is

\begin{equation}
	r = \frac{a_{\mathrm{in}}(1-e_{\mathrm{in}}^2)}{1 + e_{\mathrm{in}}\cos\nu_0},
\end{equation}
where the semi-major axis is related to the pericenter distance through
\begin{equation}
	a_{\mathrm{in}} = \frac{r_{p,\mathrm{in}}}{1-e_{\mathrm{in}}}.
\end{equation}
The radial and tangential components of the relative velocity are given by
\begin{equation}
	v_r = \sqrt{\frac{G(M_1+M_2)}{a_{\mathrm{in}}}}
	\frac{e_{\mathrm{in}}\sin\nu_0}
	{\sqrt{1-e_{\mathrm{in}}^2}},
	\qquad
	v_t = \sqrt{\frac{G(M_1+M_2)}{a_{\mathrm{in}}}}
	\frac{1+e_{\mathrm{in}}\cos\nu_0}
	{\sqrt{1-e_{\mathrm{in}}^2}}.
\end{equation}

The relative position and velocity vectors are then constructed in the orbital plane as

\begin{equation}
	\mathbf{r}_{\rm rel} = (r\cos\nu_0,\, r\sin\nu_0,\, 0)~,~~
	\mathbf{v}_{\rm rel} = (v_r\cos\nu_0 - v_t\sin\nu_0,\, v_r\sin\nu_0 + v_t\cos\nu_0,\, 0).
\end{equation}

The individual stellar positions and velocities are finally obtained in the CM frame through

\begin{equation}
	\mathbf{r}_1 = \frac{M_2}{M_1+M_2}\mathbf{r}_{\rm rel},
	\qquad
	\mathbf{r}_2 = -\frac{M_1}{M_1+M_2}\mathbf{r}_{\rm rel}~,~
	\mathbf{v}_1 = \frac{M_2}{M_1+M_2}\mathbf{v}_{\rm rel},
	\qquad
	\mathbf{v}_2 = -\frac{M_1}{M_1+M_2}\mathbf{v}_{\rm rel}.
\end{equation}

In the present simulations the binary is initialized at apocenter, corresponding to a true anomaly $\nu_0 = \pi$. Under this
condition the initial binary separation becomes $r = r_{\rm a,in} = a_{\mathrm{in}}(1+e_{\mathrm{in}})$, where $r_{\rm a,in}$ is
the apocenter distance of the inner orbit, the radial velocity vanishes ($v_r = 0$), and the initial motion is purely tangential.

\begin{figure}[h]
	\centering
	\includegraphics[width=\linewidth]{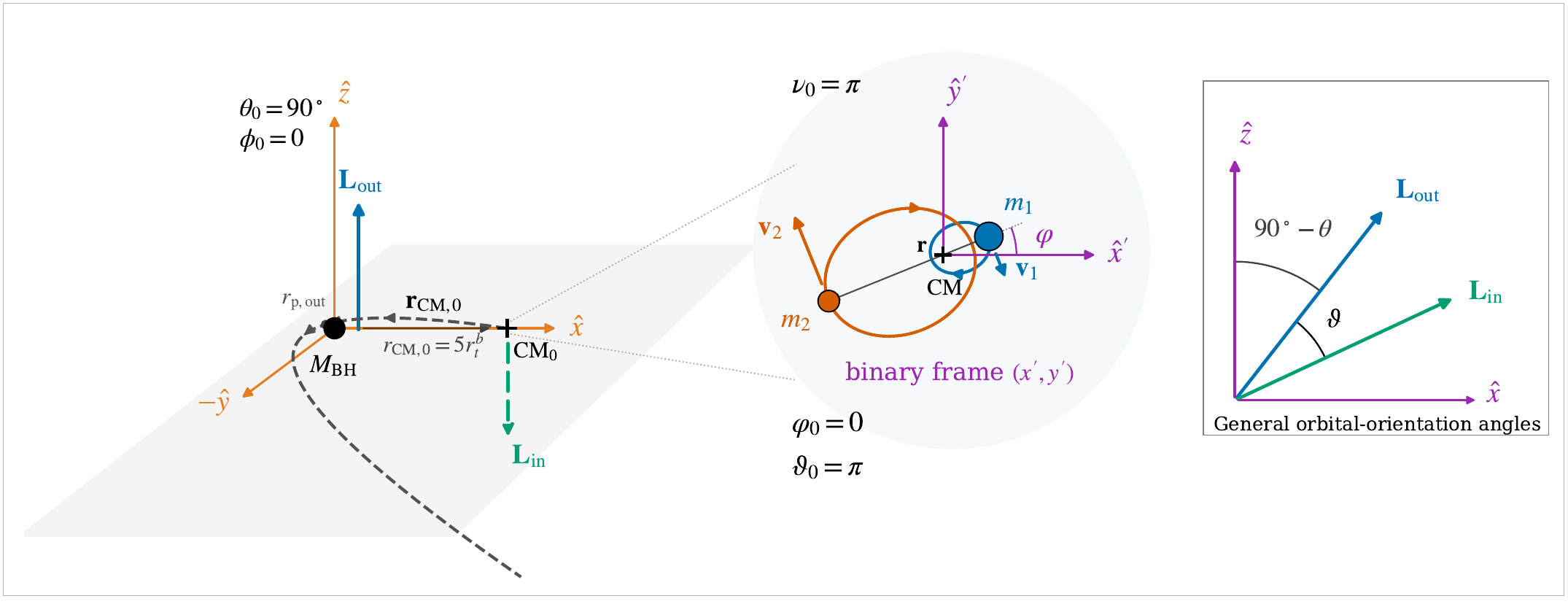}
	\caption{ Schematic illustration of the orbital configuration of an eccentric binary around a BH of mass $M_{\rm BH}$.
	\textbf{Left:} the binary CM starts at $r_{\rm CM,0}=5r_t^b$ on a parabolic outer orbit, with $\theta_0=90^\circ$ and
	$\phi_0=0^\circ$. The outer pericenter distance is $r_{\rm p,out}$, and $\mathbf{L}_{\rm out}$ and $\mathbf{L}_{\rm in}$ are the
	outer and inner orbital angular momenta. \textbf{Middle:} the inner binary in the $(x',y')$ frame, starting at apocenter
	($\nu_0=\pi$), with the component masses, separation vector $\mathbf{r}$, tangential velocities $\mathbf{v}_1$ and $\mathbf{v}_2$,
	and in-plane orientation angle $\varphi$. \textbf{Right:} orientation of $\mathbf{L}_{\rm out}$ and $\mathbf{L}_{\rm in}$ relative
	to the reference axes, with $\vartheta$ the angle between them.
	}
	\label{fig:setup}
\end{figure}

\subsection{Black Hole Background}\label{subsec:BH back}
The binary evolves in the external gravitational field of a central IMBH of mass $M_{\rm BH}$. We model the IMBH as a
Schwarzschild BH, with metric
\begin{equation}
	ds^2 =
	-\left(1-\frac{r_s}{r}\right)c^2dt^2
	+\left(1-\frac{r_s}{r}\right)^{-1}dr^2
	+r^2\left(d\theta^2+\sin^2\theta\,d\phi^2\right),
	\label{eq:schwarzschild_metric}
\end{equation}
where $r_s=\tfrac{2GM_{\rm BH}}{c^2}$ is the Schwarzschild radius. The equations of motion are obtained from geodesics in
Cartesian coordinates following \citet{Banerjee2023}. The corresponding coordinate transformation is
\begin{equation}
	r=\sqrt{x^2+y^2+z^2},
	\qquad
	\theta=\cos^{-1}\left(\frac{z}{r}\right),
	\qquad
	\phi=\tan^{-1}\left(\frac{y}{x}\right).
\end{equation}
The acceleration of each SPH particle combines a relativistic contribution from the external Schwarzschild spacetime, with the
hydrodynamic pressure, artificial viscosity, and Newtonian self-gravity terms computed within the SPH formulation of
Section~\ref{sec:numerics}. The explicit Cartesian expressions for the relativistic acceleration appear in \citet{Banerjee2023}.
The strength of the tidal interaction between the IMBH and the binary is characterized by the binary tidal radius
\citep{Sari2010},
\begin{equation}
	r_t^b =
	a_{\rm in}
	\left(\frac{M_{\rm BH}}{M_{\rm tot}}\right)^{1/3},
\end{equation}
where $a_{\rm in}$ is the initial semi-major axis of the stellar binary and $M_{\rm tot}=M_1+M_2$ is the total binary mass. We
define the binary penetration factor as
\begin{equation}
	\beta^b =
	\frac{r_t^b}{r_{p,\mathrm{out}}},
	\label{eq:beta}
\end{equation}
where $r_{p,\mathrm{out}}$ is the pericenter distance of the binary CM orbit about the IMBH. The IMBH is placed at the
origin of the coordinate system, with the binary CM position specified by the polar angle $\theta$ and azimuthal angle $\phi$. For
the Schwarzschild models considered here, the initial CM position is chosen in the equatorial plane, with $\theta_0=\pi/2$ and
$\phi_0=0$. The binary CM is initially placed at a distance $5r_t^b$ from the IMBH, as depicted in Figure \ref{fig:setup}. The
corresponding initial velocity of the binary CM is obtained from the conserved orbital constants of motion for geodesic motion in
the Schwarzschild spacetime. Scaling the initial position with the binary tidal radius ensures that the different binary models
begin at the same dimensionless distance from the characteristic tidal scale, while keeping the initial tidal perturbation
sufficiently weak that the binary is effectively unperturbed at the start of the simulation.

\subsection{Model Parameters and Initial Conditions}\label{subsec:initparam}

The binary tidal radius is fixed at $r_t^b\simeq30\,r_g$, where
$r_g=GM_{\rm BH}/c^2$ is the gravitational radius of the IMBH.
The binary CM is placed on a parabolic outer orbit,
with $e_{\rm out}=1$. We consider two encounter strengths,
$\beta^b=0.5$ and $1.5$, which, through Equation~\ref{eq:beta}, correspond to outer pericenter
distances of $r_{p,\rm out}=60\,r_g$ and $20\,r_g$, respectively.
These configurations therefore represent relatively shallow and
deep tidal encounters with the IMBH. The IMBH mass is not treated
as an independent parameter, but is determined by the adopted
binary tidal radius and the properties of the corresponding WD
binary. Consequently, the combinations of $M_{\rm BH}$,
$r_{p,\rm out}$, and $\beta^b$ listed in
Table~\ref{tab:parameters} define the specific encounter
configurations considered in this work. These choices allow us to
compare the hydrodynamic evolution of eccentric WD binaries subject
to different strengths of the external tidal perturbation.

We specify the orientation of the inner binary relative to its CM orbit about the IMBH by the angles $\vartheta$ and
$\varphi$, where $\vartheta_0$ is the initial angle between the angular momentum vectors of the inner binary and the outer orbit,
\begin{equation}
	\vartheta_0 =
	\cos^{-1}\left(
	\frac{\mathbf{L}_{\rm in,0}\cdot\mathbf{L}_{\rm out,0}}
	{|\mathbf{L}_{\rm in,0}|\,|\mathbf{L}_{\rm out,0}|}
	\right),
	\label{eq:binary_orientation}
\end{equation}
and $\varphi_0$ specifies the corresponding initial azimuthal orientation. We adopt $\vartheta_0=180^\circ$, corresponding to a retrograde
configuration. Following \citet{Binary2}, we focus on the retrograde configuration because such encounters generally preserve the
bound nature of the binary, whereas prograde encounters predominantly lead to binary separation through the Hills mechanism. The
prograde outcome nevertheless depends on the initial azimuthal phase, and some configurations can remain bound. Since our aim is
to investigate the subsequent mass transfer and merger of a bound binary in the IMBH tidal field, we restrict the present study to
the retrograde configuration.

\begin{table}[h]
	\centering
	\caption{Initial physical parameters adopted for the eccentric WD binary
		simulations in the presence of an IMBH. The entries correspond one to one
		from left to right for $q$, $a_{\rm in}$, and $M_{\mathrm{BH}}$, giving
		four binary configurations; each was run at both values of $\beta^b$,
		for eight simulations in total.}
	\label{tab:parameters}
	\begin{tabular}{ll}
		\hline
		\hline
		\textbf{Parameter} & \textbf{Value} \\
		\hline
		Primary WD Mass & $M_1 = 0.5\,M_{\odot}$ \\
		Mass Ratio & $q = M_2/M_1 \in \{0.2,\,0.3,\,0.4,\,0.7\}$ \\
		Inner Orbital Eccentricity & $e_{\mathrm{in}} = 0.6$ \\
		Inner Semi-Major Axis & $a_{\rm in} \in \{0.2447 \, R_{\odot},\,0.1896\, R_{\odot},\, 0.1579\, R_{\odot},\,0.1092\, R_{\odot}\}$ \\
		Outer Orbital Eccentricity & $e_{\mathrm{out}} = 1.0$  \\
		Binary Penetration Factor & $\beta^b \in \{0.5,\,1.5\}$ \\
		Outer Pericenter Distance & $r_{p,\mathrm{out}} \simeq \{60\,r_g,\,20\,r_g\}$ \\
		IMBH Mass & $M_{\mathrm{BH}} \in \{3.1\times10^5\,M_{\odot},\,2\times10^5\,M_{\odot},\,1.45\times10^5\,M_{\odot},\,7.6\times10^4\,M_{\odot}\}$ \\
		\hline
	\end{tabular}
\end{table}

The remaining initial binary parameters, $e_{\rm in}$ and $\varphi_0$,
are chosen by surveying 32,851 configurations of the $q=0.7$ system
over the full $e_{\rm in}$--$\varphi_0$ parameter space. For each
configuration, we evaluate the relative specific orbital energy of the
two stars after the IMBH encounter as
$\epsilon_b=\frac{1}{2}v_{12}^{2}-G(M_1+M_2)/r_{12}$, where $v_{12}$
and $r_{12}$ are their relative velocity and separation, respectively,
and normalize it by $\epsilon_\odot=GM_\odot/R_\odot$. The survey is
performed for $M_{\rm BH}=7.6\times10^4\,M_\odot$, corresponding to the
$q=0.7$ models in Table~\ref{tab:parameters}.

\begin{figure}[h]
	\centering
	\includegraphics[width=0.75\linewidth]{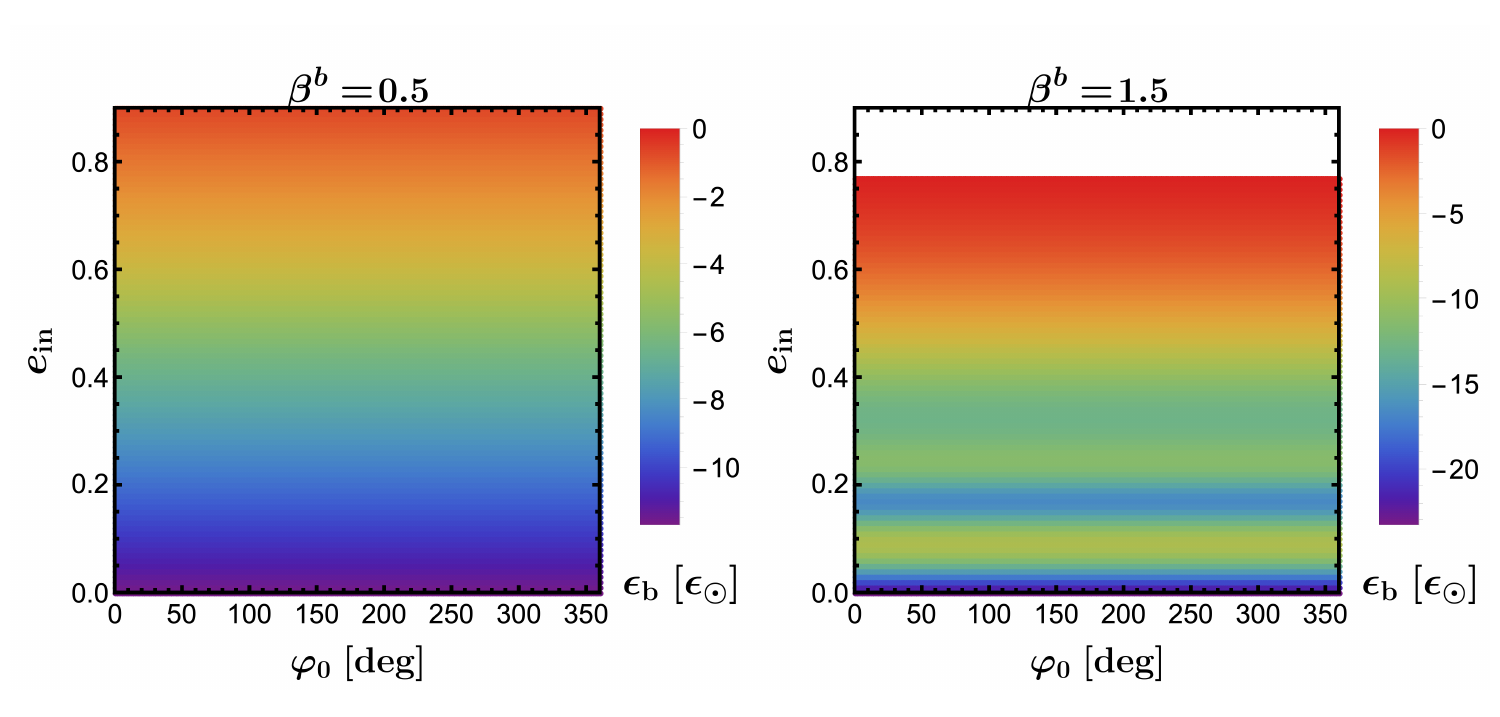}
	\caption{Specific binding energy of the retrograde binaries after pericenter passage around the IMBH, for 32,851 initial configurations with
	$q=0.7$ spanning the full $(e_{\rm in},\varphi_0)$ parameter space. \textbf{Left:} $\beta^b=0.5$. \textbf{Right:} $\beta^b=1.5$.}
	\label{binding_energy}
\end{figure}

As shown in Figure~\ref{binding_energy}, the post-encounter binding
energy depends almost entirely on the initial eccentricity and hardly
depends on $\varphi_0$ over the sampled grid. For
$\beta^b=0.5$, all configurations remain bound after the encounter, with
$-12\,\epsilon_\odot\lesssim\epsilon_b<0$. For the deeper
$\beta^b=1.5$ encounter, 28,518 configurations remain bound, spanning
approximately $-23\,\epsilon_\odot\lesssim\epsilon_b<0$, while the
remaining systems become unbound, as indicated by the white region at
large $e_{\rm in}$. Lower-eccentricity binaries remain more tightly
bound and, owing to their shorter internal orbital timescales, can
evolve significantly before experiencing the strongest
BH tidal interaction. Conversely, higher-eccentricity binaries are
more weakly bound and become more easily separated in the deeper encounter. We therefore adopt
$e_{\rm in}=0.6$ as an intermediate configuration that remains bound
while avoiding rapid pre-encounter evolution. Since the post-encounter
binding energy is effectively independent of $\varphi_0$ at fixed
$e_{\rm in}$ over the sampled grid, we set $\varphi_0=0^\circ$ for the
hydrodynamic simulations.

The primary WD is modeled as a carbon oxygen (CO) WD with mass $M_1=0.5\,M_\odot$, while the secondary is modeled as a helium (He)
WD for all cases with $M_2<0.45\,M_\odot$. Although the two WDs differ in chemical composition, both have approximately the same
mean molecular weight per electron, $\mu_e\simeq2$. Consequently, their dominant electron degeneracy pressure is described by
essentially the same Chandrasekhar relation, while the additional thermal and radiation pressure contributions are treated
consistently during the subsequent evolution. The different masses provide the desired range of binary mass ratios without
introducing a large difference in the underlying degeneracy equation of state.

The timing of the IMBH encounter relative to the inner orbit is measured by the ratio $T_p/P_0$,
where $T_p$ is the time taken by the binary CM to travel from its initial position to the outer pericenter, and $P_0$ is the
initial half-period of the inner binary. Since all systems are initialized at $5r_t^b$, this ratio provides a measure of the
number of internal binary timescales that elapse before the strongest tidal interaction with the IMBH. For an initially parabolic
outer orbit, a Newtonian estimate of the travel time from the initial position to pericenter gives
\begin{equation}
	\frac{T_p}{P_0}
	=
	\frac{\sqrt{2}}{\pi}(\beta^b)^{-3/2}
	\left(D+\frac{1}{3}D^3\right),
	\qquad
	D=\sqrt{f\beta^b-1},
	\label{eq:orbital_phasing}
\end{equation}
where $f=5$ denotes the initial CM distance in units of the binary tidal radius. Upon substituting the definitions of the binary
tidal radius and the inner binary period, the dependence on the binary mass and semi-major axis cancels, leaving $T_p/P_0$ as a
function of $\beta^b$ alone. Thus, for fixed $\beta^b$, all mass ratios have the same Newtonian value of $T_p/P_0$. 
Newtonian estimates of the orbital parameters appear in Table \ref{tab:orbital_phasing}.

\begin{table}[h]
	\centering
	\caption{Newtonian estimates for the encounter timing for the WD binary models. Here $P_0$ is the initial half-period of
	the inner binary and $T_p$ is the Newtonian travel time for the binary CM to reach the outer pericenter from the initial position
	at $5r_t^b$. The values of $T_p$ and $P_0$ are given in minutes.}
	\label{tab:orbital_phasing}
	\begin{tabular}{cccccccc}
		\hline
		\hline
		$q$ & $M_1\,(M_\odot)$ & $M_2\,(M_\odot)$ & $P_0\,(\mathrm{min})$ & $T_p\,(\beta^b=0.5)$ & $T_p\,(\beta^b=1.5)$ & $T_p/P_0\,(\beta^b=0.5)$ & $T_p/P_0\,(\beta^b=1.5)$ \\
		\hline
		0.2 & 0.5 & 0.10 & 13.04 & 30.50 & 25.80 & 2.34 & 1.98 \\
		0.3 & 0.5 & 0.15 &  8.54 & 19.98 & 16.90 & 2.34 & 1.98 \\
		0.4 & 0.5 & 0.20 &  6.26 & 14.64 & 12.38 & 2.34 & 1.98 \\
		0.7 & 0.5 & 0.35 &  3.27 &  7.64 &  6.47 & 2.34 & 1.98 \\
		\hline
	\end{tabular}
\end{table}

\section{Numerical Setup}
\label{sec:numerics}
Numerical simulations are performed using the three dimensional SPH code inspired from PHANTOM \citep{Price} and described in
\citet{Banerjee2023}, with the binary-star implementation introduced in \citet{Binary1}. The code solves self-gravitating
hydrodynamics in the external relativistic tidal field of the Schwarzschild IMBH. The SPH formulation employs adaptive smoothing
lengths, Newtonian self-gravity, artificial viscosity, and artificial conductivity. The stellar fluid is described by the
Chandrasekhar equation of state for WDs, supplemented by thermal and radiation pressure contributions as described in
Section~\ref{subsec:thermo}.

Shock capturing is implemented using the time-dependent artificial viscosity prescription of \citet{MM1997}, together with the
Balsara switch \citep{Balsara1995}. We adopt $\alpha_{\rm min}^{\rm AV}=0.1$, $\alpha_{\rm max}^{\rm AV}=1.0$, and $\beta^{\rm
AV}=2.0$. The low minimum viscosity is adopted, as in \citet{Dan2011}. A Courant factor of $0.1$ is adopted to limit
numerical velocity noise and anomalous angular momentum transport, while artificial conductivity is included with $\alpha_u=0.5$.
Self-gravity is calculated using a hierarchical tree algorithm with kernel-softened near-field interactions. The neighbor-search
and gravity trees are constructed using recursive coordinate bisection, with each node split along its longest spatial dimension
until the average number of particles
per leaf drops below ten \citep{GaftonRosswog2011}. Near-field forces are evaluated by direct summation, while
distant nodes are treated using monopole and quadrupole moments with $\theta_{\rm MAC}=0.5$ (multipole acceptance criterion).
Adaptive smoothing length corrections are included consistently within the conservative SPH formulation.

The initial WD models are constructed on a hexagonal close-packed lattice and subsequently stretch-mapped to reproduce the desired
equilibrium density profiles \citep{partial26}. The stellar models are dynamically relaxed before being assembled into the
eccentric binary configuration described in Section~\ref{subsec:Binary system}. The resulting binary is then initialized on its
orbit around the IMBH according to the procedure described in Section~\ref{subsec:BH back}. This construction follows the
binary-star implementation of \citet{Binary1}, with the present work extending the setup from the circular binary configuration
considered there to an eccentric binary. The relativistic contribution from the external BH field is given by a standard
expression
\begin{equation}
	\ddot{x}^{i}_{a}
	=
	-\left(
	g^{i\lambda}
	-
	g^{0\lambda}\dot{x}^{i}_{a}
	\right)
	\left[
	\frac{\partial g_{\mu\lambda}}{\partial x^\nu}
	-\frac{1}{2}
	\frac{\partial g_{\mu\nu}}{\partial x^\lambda}
	\right]_{a}
	\dot{x}^{\mu}_{a}\dot{x}^{\nu}_{a},
	\label{eq:geodesic}
\end{equation}
where $x^\mu=(t,x^i)$ denotes the spacetime coordinates, overdots denote derivatives with respect to coordinate time $t$, and
$g_{\mu\nu}$ is the Schwarzschild metric given in Equation \ref{eq:schwarzschild_metric}.

\subsection{Hydrodynamics and Thermodynamic Modeling}
\label{subsec:thermo}
We adopt the equation of state based on \citet{Benz1989}, consisting of zero-temperature, fully relativistic degenerate electrons, an
ideal ion gas, and radiation, $p = p_{\rm deg}(\rho) + \rho k_{\rm B}T/(\mu_i m_{\rm u}) + a_{\rm rad}T^4/3$. The specific
internal energy evolves as $du/dt = -(p/\rho)\nabla\cdot\mathbf{v}$ along with artificial viscosity and conductivity terms. The
temperature is obtained by Newton Raphson iteration from $u - u_{\rm deg}(\rho) = 3k_{\rm B}T/(2\mu_i m_{\rm u}) + a_{\rm
rad}T^4/\rho$, with $\mu_i = 14$ (CO) and $4$ (He). No nuclear network is evolved. The possible dynamical importance of nuclear
burning is discussed in Appendix~\ref{nuclear_ignition}.

\subsection{Numerical Resolution and Convergence}\label{subsec:resolution}

The fiducial simulations employ $10^5$ SPH particles per WD
($N=2\times10^5$ in total). Using equal particle numbers in the two
components, rather than equal particle masses, keeps the lower-mass
donor well resolved during expansion and disruption. The particle
mass is uniform within each WD, $m_{{\rm part},k}=M_k/10^5$, with a
maximum inter-component contrast of a factor of $5$ for $q=0.2$.
Finite resolution imposes a minimum resolvable mass-transfer rate of
$\dot M_{\rm lim}\sim m_{\rm part}/P_{\rm orb}
\sim10^{-9}$--$10^{-8}\,M_\odot\,{\rm s}^{-1}$ \citep{Rosswog2026},
so only the onset of stripping from the outermost layers is sensitive
to particle sampling; the subsequent RLOF proceeds at much higher
rates. Representative runs with $3\times10^5$ particles per WD
reproduce the orbital, mass-transfer, angular-momentum, and merger
evolution, with modest differences confined mainly to the late
nonlinear phase (Appendix~\ref{app:resolution}).

\subsection{Bound Core Identification and Merger Criterion}\label{subsec:boundcore}
To track the evolution of the binary during repeated tidal interactions, we identify the self-bound material associated with each
WD throughout the simulation. The bound material associated with each WD core is identified using an iterative energy-based
procedure following \citet{Guillochon2013}, with the binding of each particle evaluated separately with respect to the two stellar
cores. Following \citet{Li2021}, we define a generalized specific binding quantity for particle $i$ with respect to core $a$ as
\begin{equation}
	h_{i,a}
	=
	\frac{1}{2}
	\left|
	\mathbf{v}_i-\mathbf{v}_{{\rm core},a}
	\right|^2
	+
	\Phi_{i,a}
	+
	u_i
	+
	\frac{p_i}{\rho_i},
	\label{eq:binding_parameter}
\end{equation}
where $\mathbf{v}_i$ is the velocity of particle $i$, $\mathbf{v}_{{\rm core},a}$ is the CM velocity of core $a$, and
$\Phi_{i,a}$ is the gravitational potential energy per unit mass of particle $i$ due to the material associated with core $a$. The
quantities $u_i$, $p_i$, and $\rho_i$ denote the specific internal energy, pressure, and density of the particle, respectively.
The $p_i/\rho_i$ term accounts for the enthalpy contribution to the binding criterion. The procedure is initialized by identifying
the particle with the maximum density associated with each WD and its velocity is used as the initial estimate of the core velocity.
The set of particles assigned to each core is then iteratively updated, and the core velocity is recalculated from the CM
of the currently bound particles,
\begin{equation}
	\mathbf{v}_{{\rm core},a}
	=
	\frac{\displaystyle\sum_{i\in a}m_i\mathbf{v}_i}
	{\displaystyle\sum_{i\in a}m_i},
	\label{eq:core_velocity}
\end{equation}
This procedure is repeated until the bound core mass converges. Each particle is assigned 
according to the following rule : a) if $h_{i,a}<0$ and $h_{i,b}>0$, it is assigned to core $a$, b)
if $h_{i,a}<0$ and $h_{i,b}<0$, the particle is assigned to the core for which the binding is stronger, e.g. to core $a$ if 
$h_{i,a}<h_{i,b}$ c)
if $h_{i,a}>0$ and $h_{i,b}>0$, it is not assigned to either core. 
Unassigned particles may still be bound to the binary as a whole, so this procedure is used to only measure core masses.
Mass transfer between the binary components is identified when particles initially associated with one WD become assigned
to the companion core.

A \textbf{merger event} is defined as the stage at which one of the two self-bound cores can no longer be identified as an
independent core, with its bound material becoming predominantly associated with the companion core. Since all models considered
here have unequal component masses ($q<1$), the two stellar components can be distinguished throughout the pre-merger evolution,
allowing the merger time to be determined from the evolution of the bound core masses and their separation.

\subsection{Energy and Angular Momentum}
\label{subsec:energetics_angmom}

To quantify the redistribution of energy and angular momentum during the
interaction, we follow the inner binary orbital angular momentum, the spin
angular momentum of the two bound cores, the orbital angular momentum of
the binary CM about the IMBH, the internal orbital energy of
the binary, and the thermal energy of the bound stellar material. The
properties of the inner binary are evaluated from the two bound cores
identified using the iterative procedure described in
Section~\ref{subsec:boundcore}. Their instantaneous CM is
$\mathbf{x}_{\rm bin}=(M_{b,1}\mathbf{x}_1+M_{b,2}\mathbf{x}_2)/
(M_{b,1}+M_{b,2})$, with an analogous expression for
$\mathbf{v}_{\rm bin}$. Here $M_{b,k}$, $\mathbf{x}_k$, and
$\mathbf{v}_k$ are the bound mass, CM position, and
CM velocity of core $k$.
The orbital angular momentum of the inner binary and the spin angular momentum of core $k$ are defined as
\begin{equation}
	\mathbf{L}_{\rm in}
	=
	\sum_{k=1}^{2}
	M_{b,k}
	\left(\mathbf{x}_k-\mathbf{x}_{\rm bin}\right)
	\times
	\left(\mathbf{v}_k-\mathbf{v}_{\rm bin}\right),
	\qquad
	\mathbf{S}_k
	=
	\sum_{i\in{\cal C}_k}
	m_i
	\left(\mathbf{x}_i-\mathbf{x}_k\right)
	\times
	\left(\mathbf{v}_i-\mathbf{v}_k\right),
	\label{eq:Lin_Sk}
\end{equation}
where ${\cal C}_k$ denotes the set of particles bound to core $k$. We denote the magnitude of the internal orbital angular momentum by
$L_{\rm in}=|\mathbf L_{\rm in}|$. The signed component $L_{{\rm in},z}$ is useful to retain information about the
orbital sense.

The orbital angular momentum of the binary CM about the IMBH
is monitored separately through
$L_{\rm out}^{\rm CM}=u_\phi=u^t g_{\phi\phi}\dot{\phi}$, with
$u^t$, $g_{\phi\phi}$, and $\dot{\phi}$ evaluated at the instantaneous
binary CM position. This quantity is used only as an indicator
of the external orbital motion.

The internal orbital energy of the bound binary is estimated as
\begin{equation}
	E_{\rm orb}
	=
	\frac{1}{2}\mu
	\left|\mathbf{v}_1-\mathbf{v}_2\right|^2
	-
	\frac{G M_{b,1}M_{b,2}}
	{\left|\mathbf{x}_1-\mathbf{x}_2\right|},
	\label{eq:Eorb}
\end{equation}
where $\mu=M_{b,1}M_{b,2}/(M_{b,1}+M_{b,2})$ is the instantaneous
reduced mass. We use $E_{\rm orb}$ to see how the inner binary is bound. It is not the conserved energy of the
full binary and IMBH system.

Finally, the thermal energy retained within the two bound stellar cores is
$U_{\rm tot}=\sum_{i\in{\cal C}_1}m_i u_i+
\sum_{i\in{\cal C}_2}m_i u_i$, where $u_i$ is the specific internal
energy of particle $i$. Material unbound from both cores is excluded from
$U_{\rm tot}$, so its evolution traces the heating of the surviving
stellar components during tidal deformation, shocks, and mass transfer.

\section{Dynamical Evolution of White Dwarf Binaries}
\label{sec:results}

Now we present our main result. First we describe evolution of isolated binaries, which serve as reference for the IMBH simulations . We then examine how the IMBH
encounter modifies the orbital dynamics, mass transfer, angular momentum redistribution, energetics, and merger evolution across
different mass ratios and encounter strengths.

\subsection{Reference Evolution of Isolated Eccentric White Dwarf Binaries}
\label{subsec:iso_evolution}

\begin{figure}[h] \epsscale{1.17} \plottwo{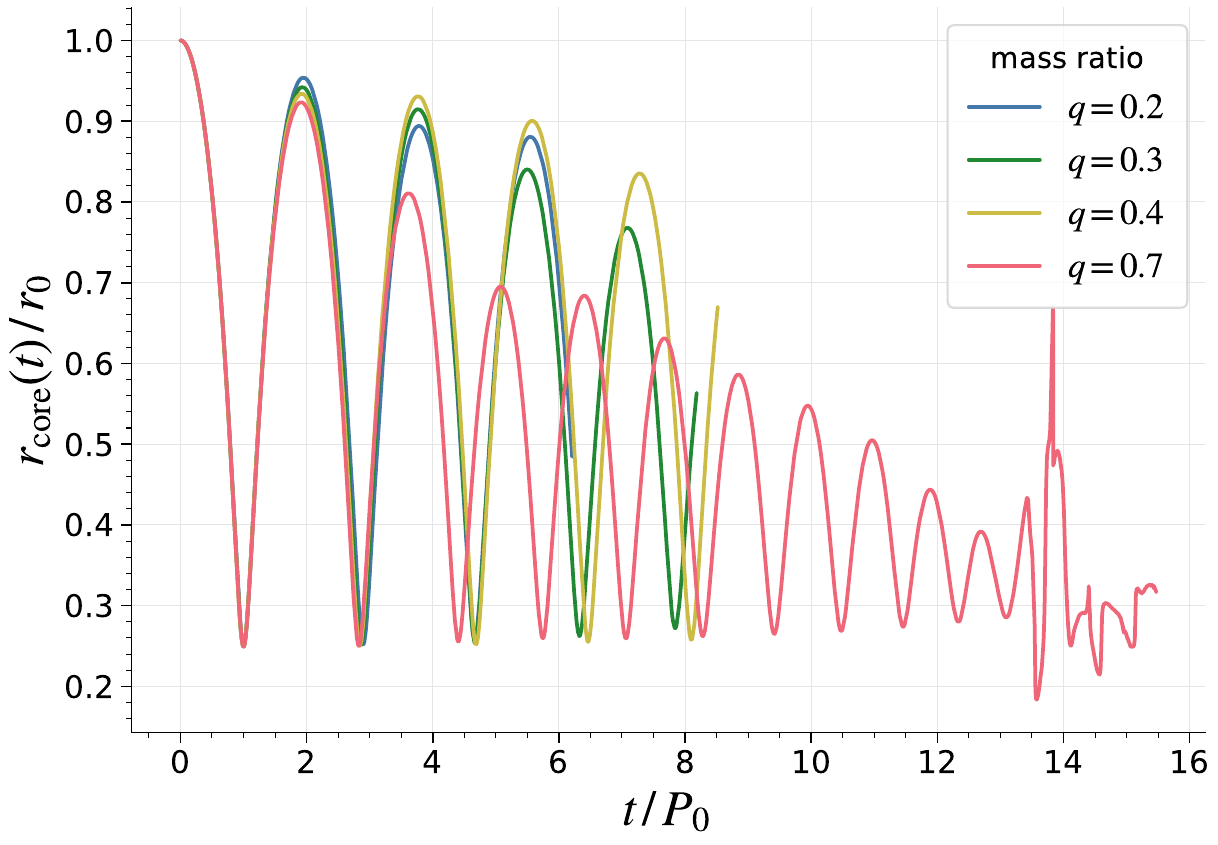}{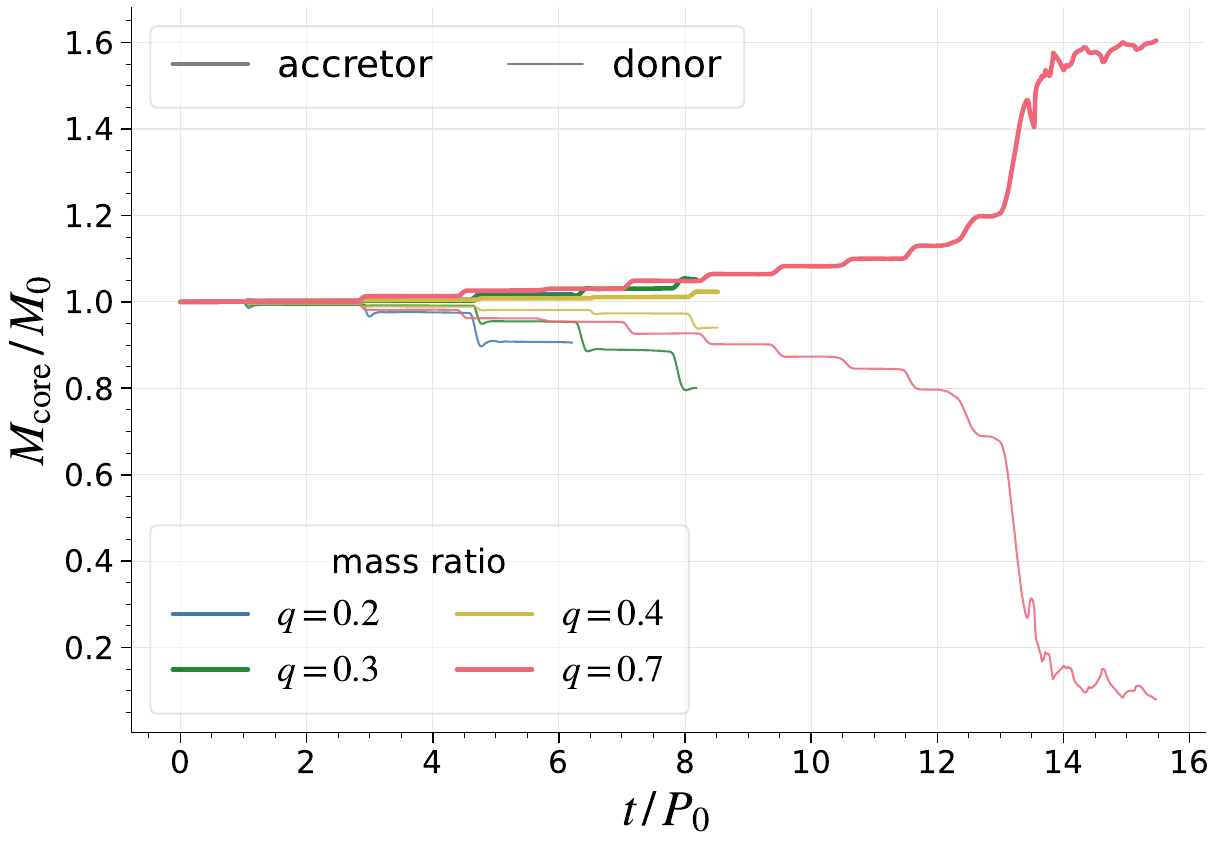}
	\caption{ Dynamical evolution of isolated eccentric WD binaries with $e_{\rm in}=0.6$. \textbf{Left:} separation of the bound core
	CMs, $r_{\rm core}/r_0$, for $q=0.2$, $0.3$, $0.4$, and $0.7$, with $r_0=0.3915$, $0.3034$, $0.2526$, and
	$0.1747\,R_\odot$. \textbf{Right:} bound core mass $M_{\rm core}$, normalized by the initial mass $M_0$ of each star, for the
	accretor (thick) and donor (thin). Time is in units of the initial binary half-period $P_0$ (Table~\ref{tab:orbital_phasing}). The
	$q=0.7$ run is followed to merger at $t/P_0\simeq15$, and the $q\leq0.4$ runs end at $t/P_0\simeq6$--$8.5$.
	}
	\label{fig:isolated_binary}

\end{figure}

Figure~\ref{fig:isolated_binary} shows the bound core separation $r_{\rm core}/r_0$ and bound core masses $M_{\rm core}/M_0$ for
the isolated binaries, which we initialized at apocenter with $e_{\rm in}=0.6$, so that the first pericenter occurs at
$t/P_0\simeq1$ with $r_{\rm core}/r_0=(1-e_{\rm in})/(1+e_{\rm in})=0.25$. With successive passages the apocenter separation
decreases, with relatively less change in the pericenter, indicating orbital contraction and decreasing eccentricity. This
evolution is gradual for $q\le0.4$ but considerably stronger for $q=0.7$.

The bound core masses in the right panel show that mass exchange is episodic, and concentrated around pericenter. Between
successive close passages the core masses remain nearly constant and changes occur during the short intervals of strongest stellar
interactions. Some passages also show temporary decreases followed by partial recovery of the identified core mass. These short
lived dips are not real mass loss, they occur as near pericenter, tidally stretched outer material briefly fails the binding
criterion and is excluded from the core, then rejoins back once the star relaxes.

The response to repeated mass transfer depends on the binary mass ratio. At the first pericenter passage, the fractional change in
the donor mass shows a systematic dependence on $q$, with the lower mass donors undergoing larger fractional changes in their
bound mass. This is expected, as lower mass WDs have larger radii and are less tightly bound. This initial ordering, however, is not
preserved during subsequent passages. Once mass transfer begins, each binary follows a different evolutionary trajectory. The
component masses, orbital separation, degree of Roche lobe overfilling, and timing of subsequent pericenter passages are
progressively modified by the preceding interactions.

For $q=0.7$ the mass exchanged per passage grows with each encounter, leading to merger at $t/P_0\simeq15$, once the transferred material has settled. At late times, the
material unbound from both stellar cores remains only a few percent of the initial total binary mass (not shown), indicating that
most of the mass removed from the donor during the runaway phase is incorporated into the accretor and subsequent merger remnant.
The lower $q$ binaries evolve more gradually and remain as distinct bound cores over the simulated intervals.

\subsection{Tidal Evolution in the Presence of an IMBH}

\subsubsection{Orbital Dynamics}
\label{subsubsec:orbsep}

\begin{figure}[h]
	\epsscale{1.17}
	\plottwo{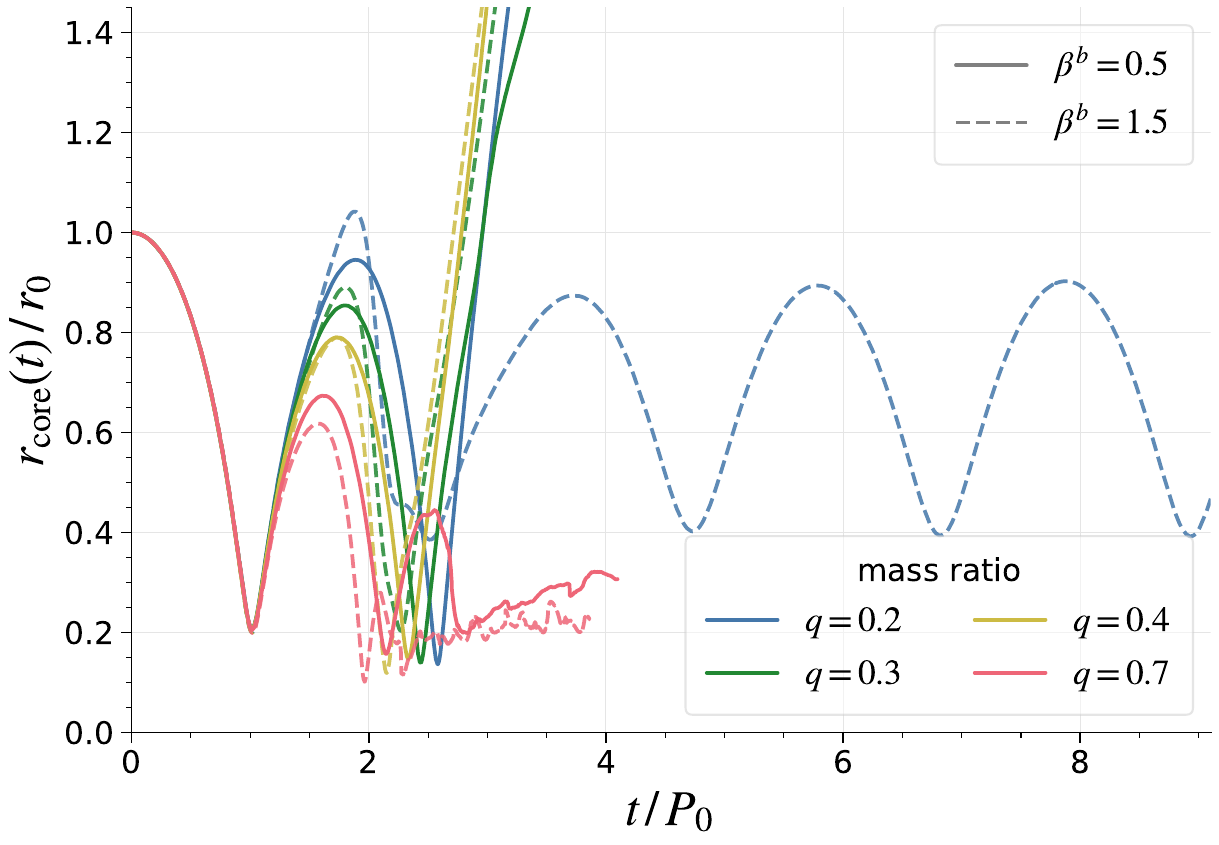}{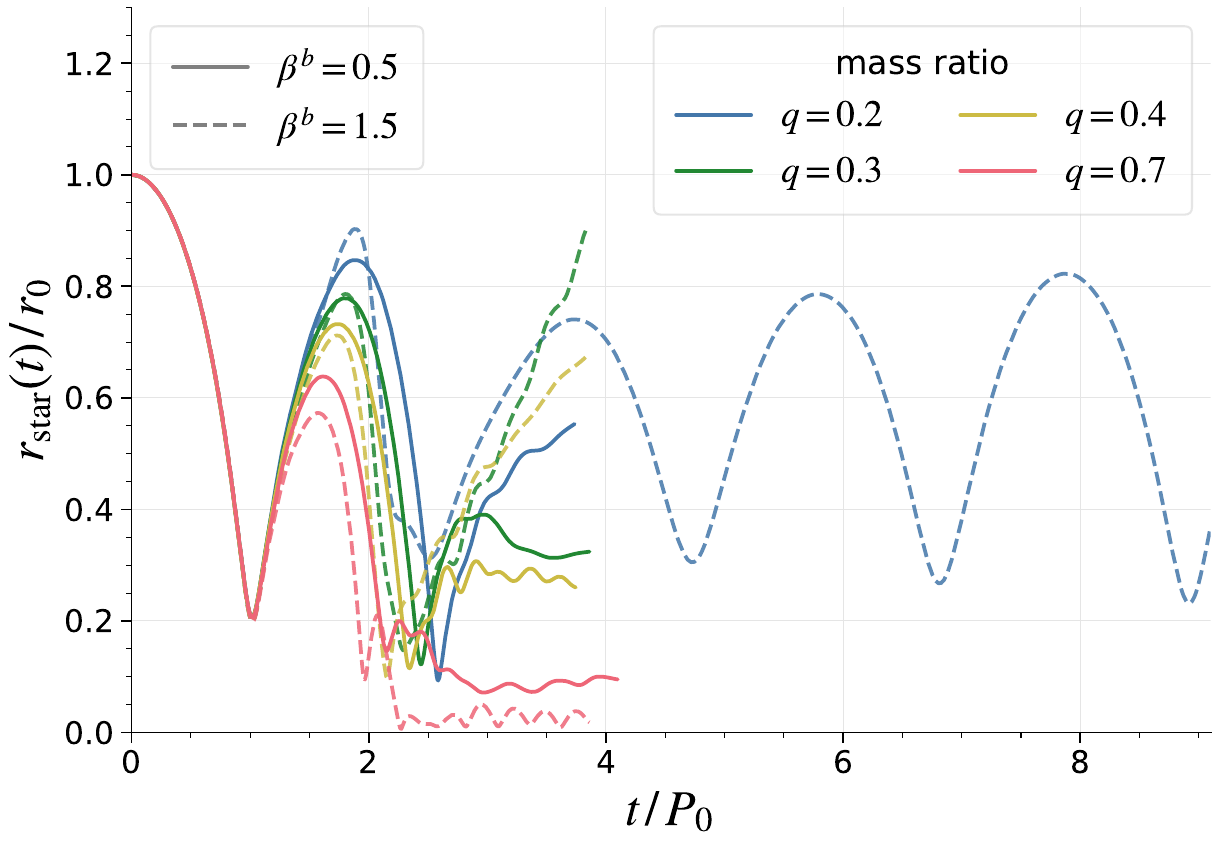}
	\caption{
		Binary separation in the IMBH encounters for $\beta^b=0.5$ (solid)
		and $1.5$ (dashed), with $r_0$ as in
		Figure~\ref{fig:isolated_binary}. \textbf{Left:} bound core
		CM separation, $r_{\rm core}/r_0$. \textbf{Right:}
		separation of the stellar density maxima, $r_{\rm star}/r_0$.
		Time is in units of $P_0$. Runs terminate shortly after merger,
		except for the surviving $q=0.2$, $\beta^b=1.5$ model, which is
		followed for longer.
	}
	\label{fig:core_separation}
\end{figure}

To characterize the dynamical response of the binary to the IMBH tidal
field, Figure~\ref{fig:core_separation} compares the separation of the
bound-core CMs, $r_{\rm core}$, with that of the stellar density maxima,
$r_{\rm star}$. Throughout Figures~\ref{fig:core_separation}--\ref{fig:energy},
the runs terminate shortly after merger, except for the
$q=0.2$, $\beta^b=1.5$ model, which retains two bound cores throughout
the simulated interval and is followed for longer. We refer to this
configuration below as the surviving model.

\begin{table}[h]
	\centering
	\caption{
		Measured encounter timing for the WD binary models. $P_0$ is the
		initial Newtonian half-period of the inner binary, $T_p$ is the
		measured time at which the binary CM reaches the outer pericenter,
		and $t_{\rm in,2}$ is the measured time of the second inner
		pericenter passage. Comparing $T_p/P_0$ and
		$t_{\rm in,2}/P_0$ shows where each binary is in its inner orbit when
		it reaches outer pericenter. 
	}
	\label{tab:actual_orbital_phasing}
	\begin{tabular}{cccccc}
		\hline
		\hline
		$q$ & $P_0\,(\mathrm{min})$ &
		$T_p/P_0\,(\beta^b=0.5)$ &
		$T_p/P_0\,(\beta^b=1.5)$ &
		$t_{\rm in,2}/P_0\,(\beta^b=0.5)$ &
		$t_{\rm in,2}/P_0\,(\beta^b=1.5)$ \\
		\hline
		0.2 & 13.04 & 2.42 & 2.08 & 2.63 & 2.54 \\
		0.3 &  8.54 & 2.42 & 2.08 & 2.45 & 2.29 \\
		0.4 &  6.26 & 2.42 & 2.08 & 2.35 & 2.16 \\
		0.7 &  3.27 & 2.42 & 2.08 & 2.15 & 1.96 \\
		\hline
	\end{tabular}
\end{table}

At fixed encounter strength, $T_p/P_0$ is nearly independent of mass
ratio, with $T_p/P_0\simeq2.42$ for $\beta^b=0.5$ and $\simeq2.08$
for $\beta^b=1.5$, as expected from the scaled construction of
Section~\ref{sec:setup}. The measured values are a few percent larger
than the corresponding Newtonian estimates, reflecting the difference
between the idealized construction and the actual trajectory in the
Schwarzschild background. In contrast, the second inner pericenter occurs
progressively later for decreasing $q$. At outer pericenter, only the
$q=0.7$ binary has already passed its second inner pericenter for
$\beta^b=1.5$, whereas for $\beta^b=0.5$ this is true for
$q=0.4$ and $0.7$. The strongest tidal perturbation therefore acts at
different phases of the inner orbit across the models.

The first inner passage produces only modest differences among the IMBH
encounters, whereas the outcomes diverge strongly during the second cycle.
All configurations except $q=0.2$, $\beta^b=1.5$ merge during or shortly
after this passage, compared with $t/P_0\simeq15$ for the isolated
$q=0.7$ binary. For $q=0.2$ at $\beta^b=0.5$ and for the
$q=0.3$ and $0.4$ models, merger proceeds through disruption of the donor
core, while the $q=0.7$ systems undergo rapid coalescence during the same
strongly perturbed cycle.

The $q=0.2$, $\beta^b=1.5$ model instead retains two bound cores.
After the encounter, $r_{\rm core}$ oscillates between
$r_{\rm p,core}/r_0\simeq0.39$ and
$r_{\rm a,core}/r_0\simeq0.89$. For a Keplerian orbit, this gives
$a_{\rm in}\simeq1.02\,a_{\rm in,0}$ and $e_{\rm in}\simeq0.39$. The orbital period and the energy 
and angular momentum give similar values,
$a_{\rm in}\simeq1.02$--$1.03\,a_{\rm in,0}$ and
$e_{\rm in}\simeq0.37$--$0.39$. The encounter thus 
leaves the semi-major axis almost unchanged but substantially lowers the eccentricity
from $0.6$ to about $0.4$. Since its $\beta^b=0.5$
counterpart merges, so a deeper encounter does not always lead to merger, and the
inner orbital phase at outer pericenter plays a crucial role. 

The two separation measures serve different purposes. While both cores
remain compact, $r_{\rm core}$ provides the more direct measure of their
relative orbit. During strong stripping, however, extended material still
bound to a distorted core can displace its CM, whereas the
density maxima more directly trace the locations of the compact stellar
concentrations. 
In the surviving $q=0.2$, $\beta^b=1.5$
model the growing variations of $r_{\rm star}$ do not appear in
$r_{\rm core}$, the orbital period, or the energy and angular momentum. 
They therefore come from changes in internal density,
not from the binary orbit. 
We consequently use the bound core criterion
of Section~\ref{subsec:boundcore} to define merger and avoid interpreting
either separation measure as a binary orbit once a bound core is unidentifiable.

\subsubsection{Hydrodynamic Morphology}
\begin{figure}[h]
	\centering

	\gridline{
		\fig{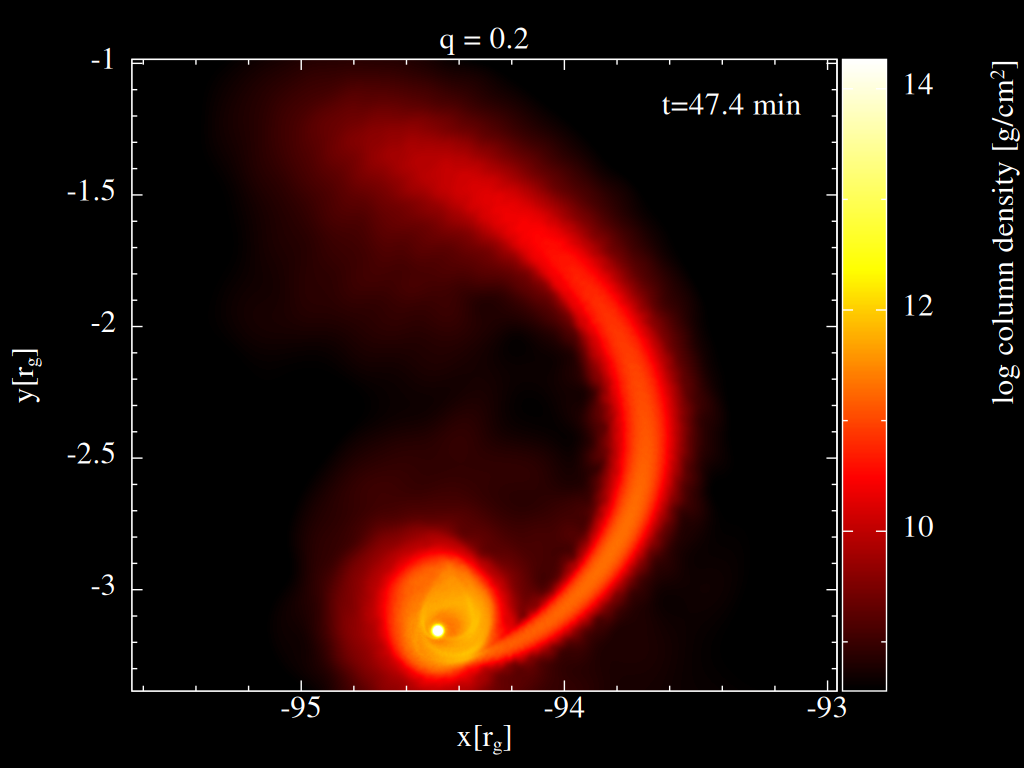}{0.50\textwidth}{}
		\fig{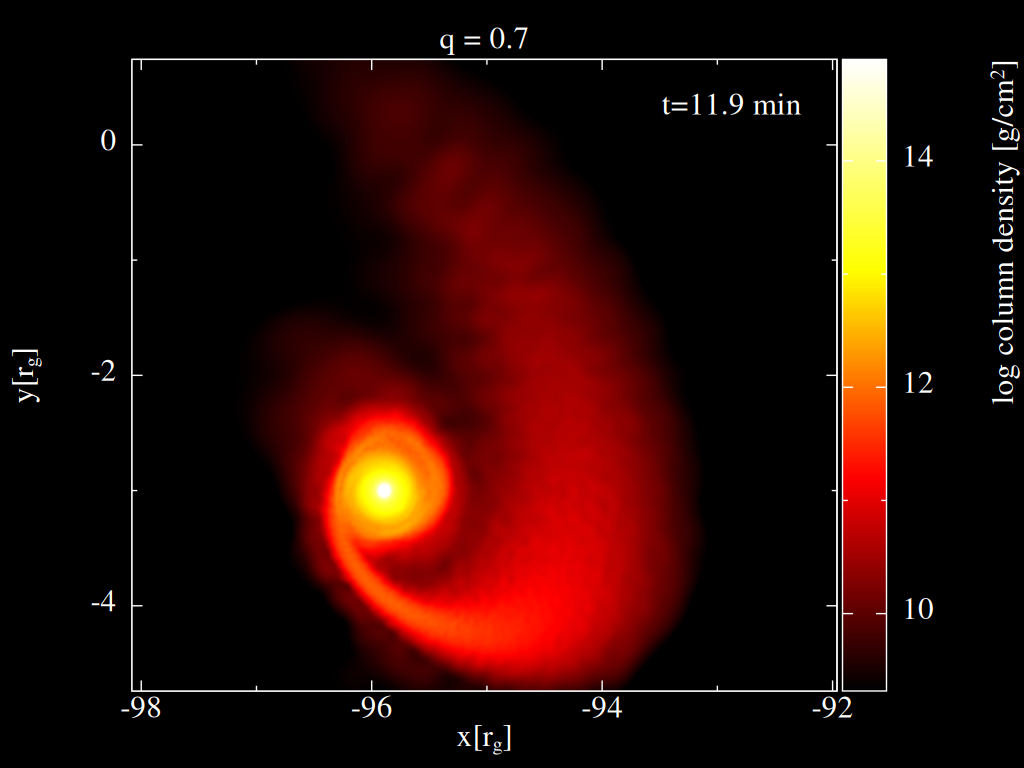}{0.50\textwidth}{}
	}
	\vspace{-0.9cm}
	\gridline{
		\fig{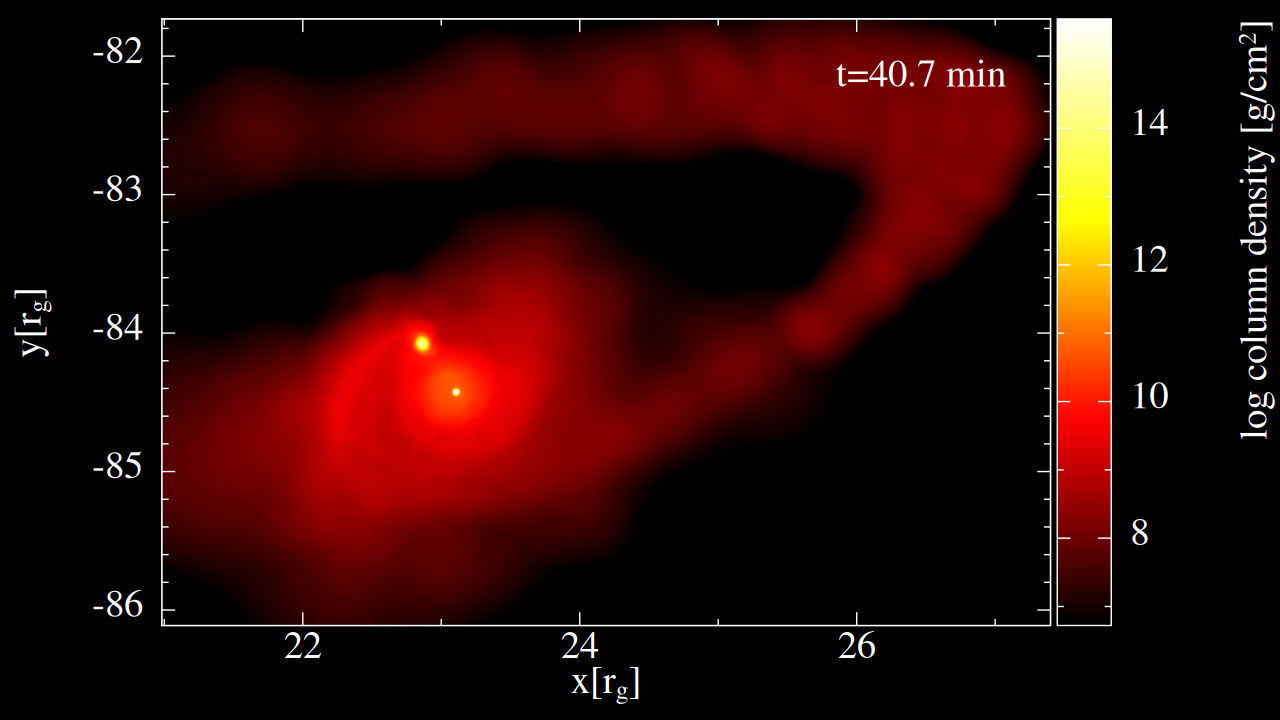}{0.50\textwidth}{}
		\fig{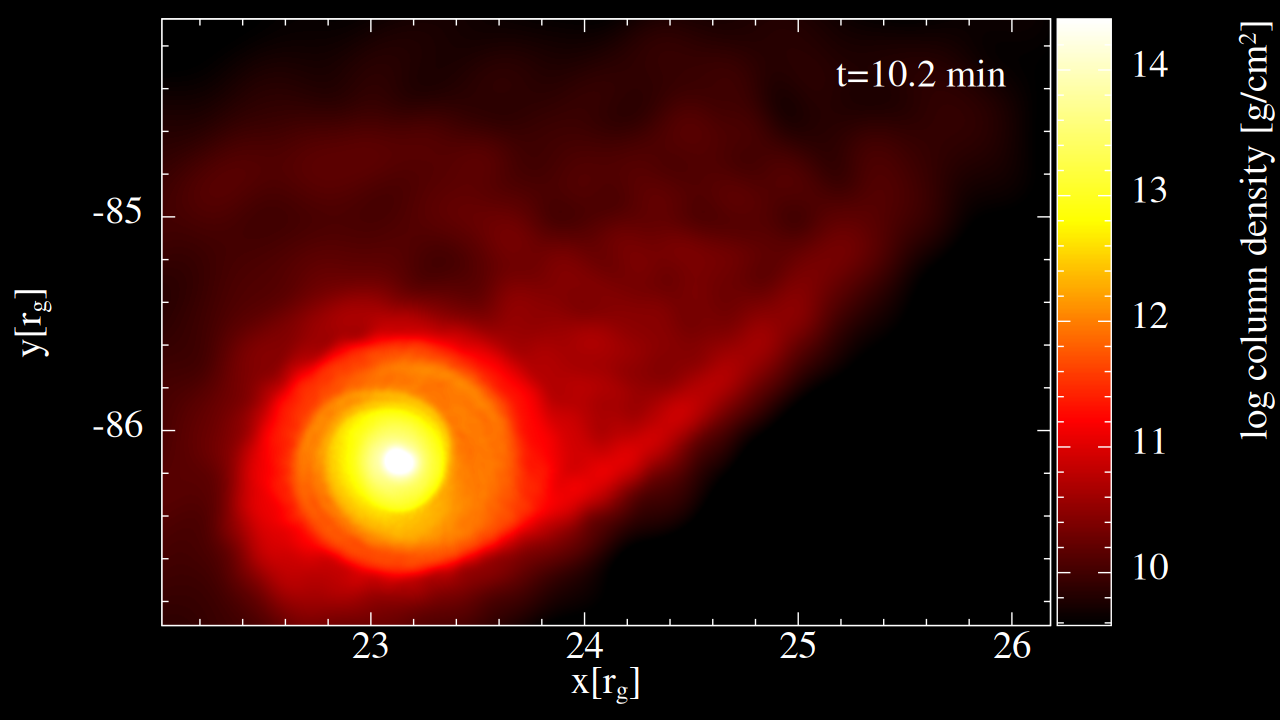}{0.50\textwidth}{}
	}
	\caption{ SPLASH column-density maps projected onto the orbital plane at $t/T_p\simeq1.5$, after the binary CM has passed the
	outer pericenter, for $q=0.2$ (left) and $q=0.7$ (right) with $\beta^b=0.5$ (top) and $1.5$ (bottom). Physical times, which differ
	between models because $T_p$ varies, are given in each panel, and coordinates are in units of $r_g$. Only the $q=0.2$,
	$\beta^b=1.5$ system (bottom left) retains two density concentrations. The other three have merged and show a single remnant
	surrounded by tidal debris.
	}
	\label{fig:snapshots}
\end{figure}

Figure~\ref{fig:snapshots} shows column-density maps for $q=0.2$ and $0.7$ at $t/T_p\simeq1.5$, after the binary CM has passed the
IMBH pericenter. The morphology reflects the outcomes of Section~\ref{subsubsec:orbsep}. Only the surviving $q=0.2$, $\beta^b=1.5$
model retains two compact density concentrations. The merged models show a single remnant surrounded by stellar debris, which
forms a long, narrow tidal stream for $q=0.2$, $\beta^b=0.5$ and broader, more diffuse structure for $q=0.7$. Density enhancements
within this debris can be identified as secondary maxima, which explains the finite late-time values of $r_{\rm star}$ after
merger.

\subsubsection{Mass Transfer and Unbound Material}
\label{subsubsec:massdynamics}

\begin{figure}[h]
	\epsscale{1.17}
	\plottwo{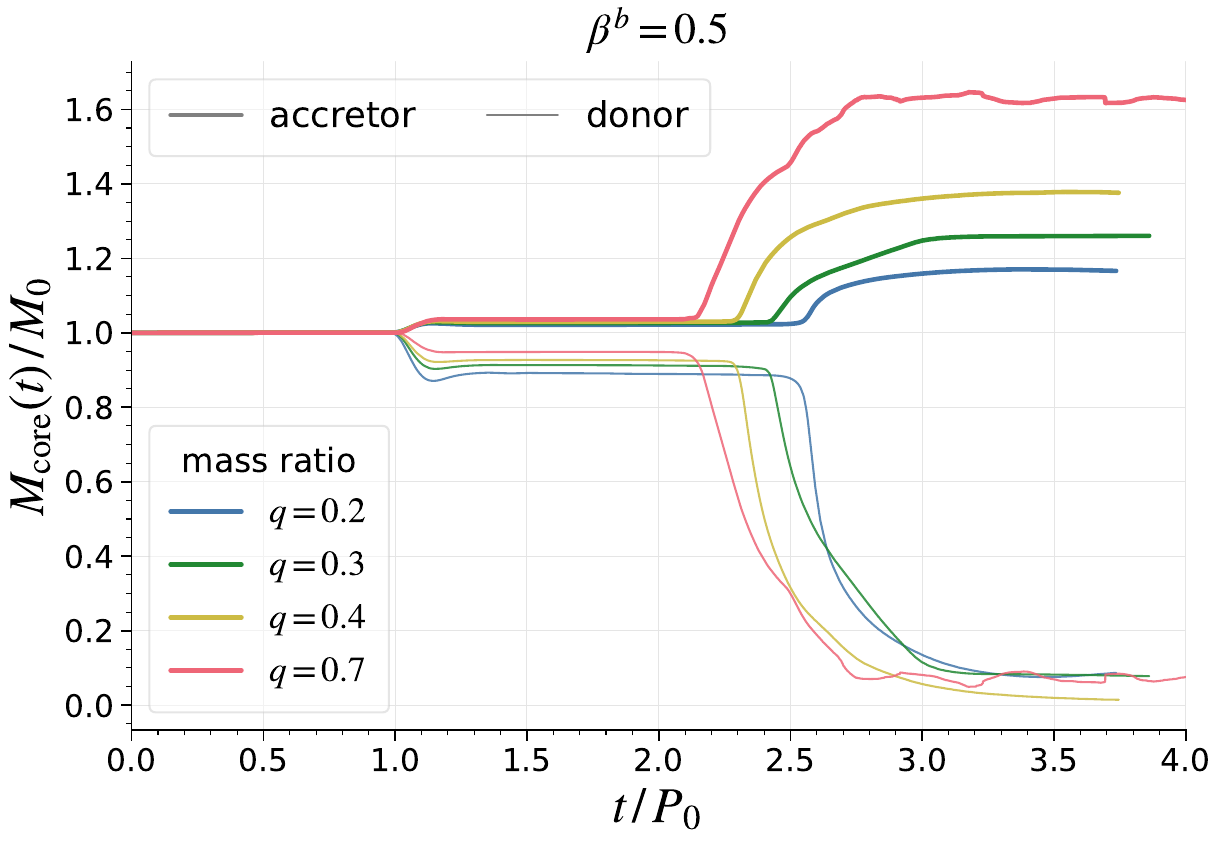}{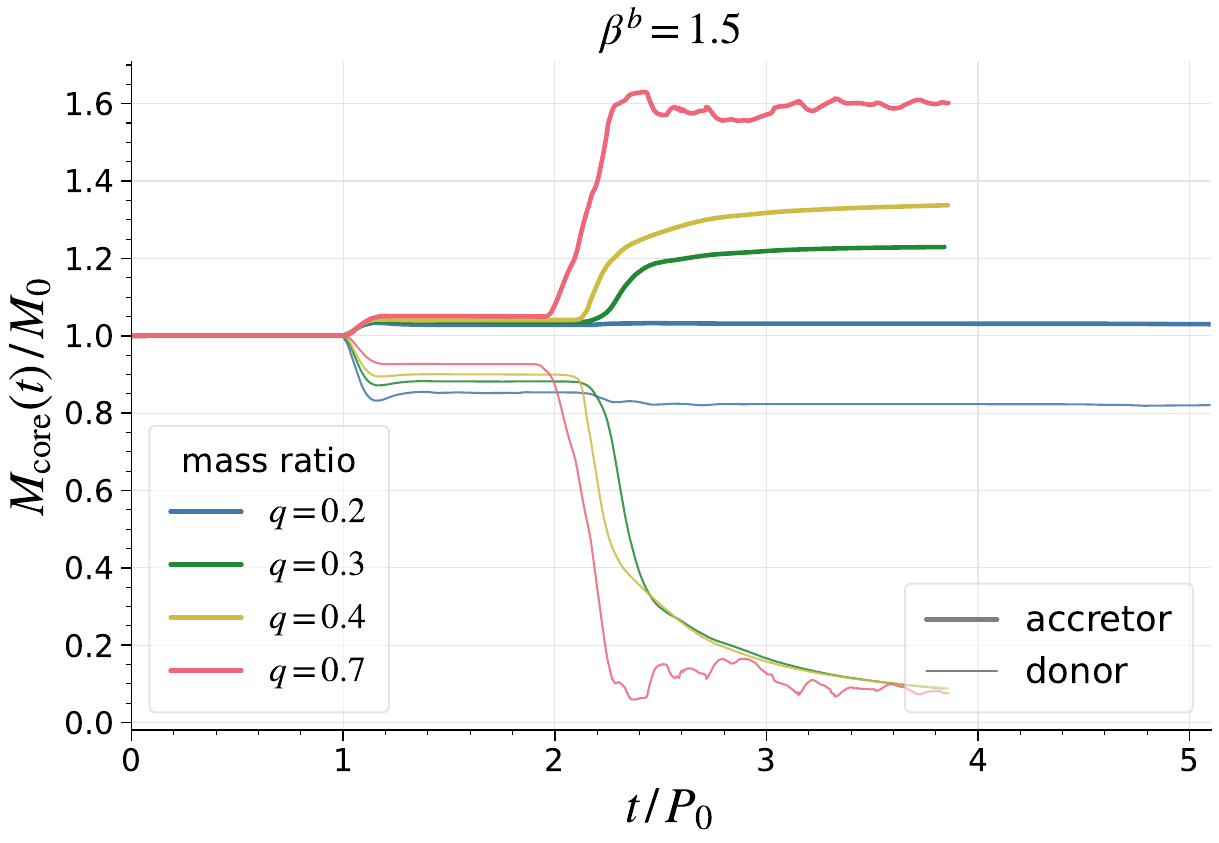}
	\caption{
		Bound core mass $M_{\rm core}$, normalized by the initial mass $M_0$
		of each star, for the accretor (thick) and donor (thin).
		\textbf{Left:} $\beta^b=0.5$. \textbf{Right:} $\beta^b=1.5$.
		Time is in units of $P_0$.
	}
	\label{fig:core_mass}
\end{figure}

\begin{figure}[h]
	\epsscale{0.6}
	\plotone{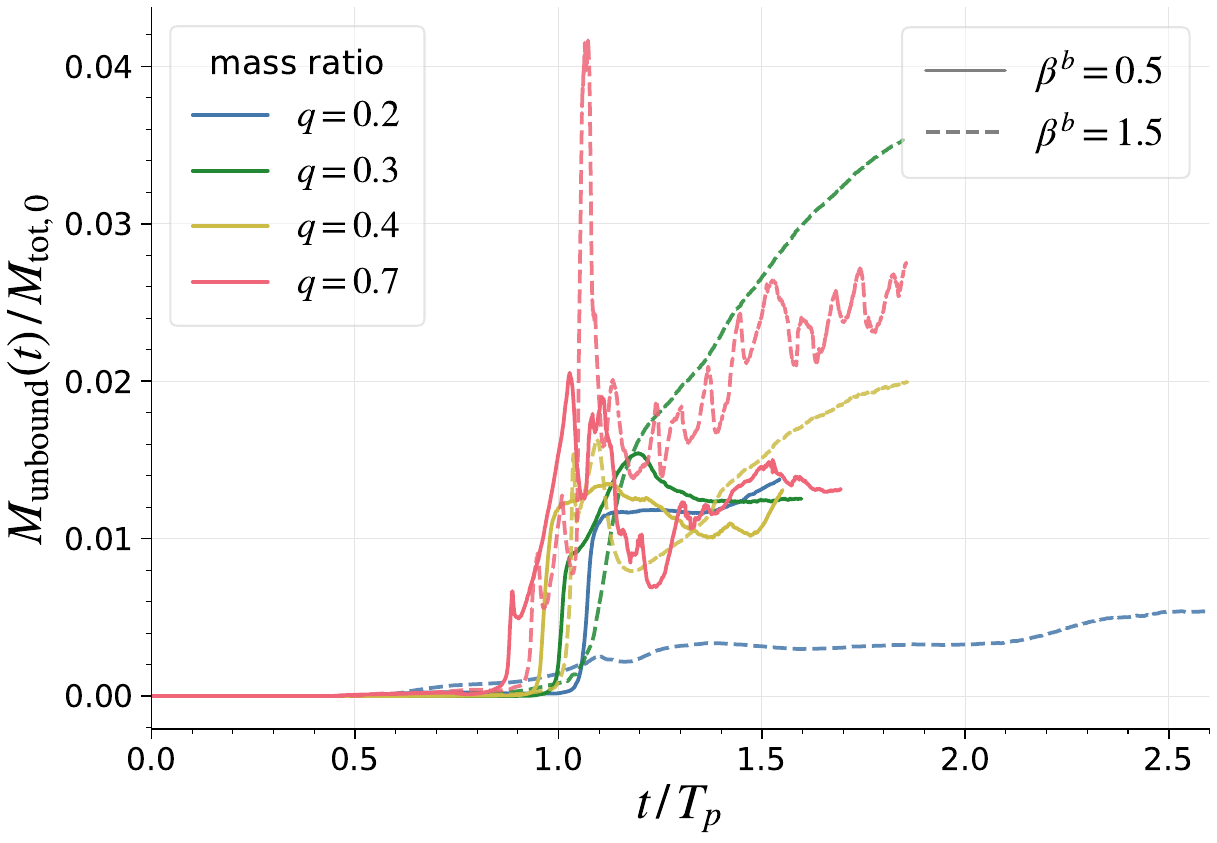}
	\caption{
		Instantaneous mass unbound from both stellar cores,
		$M_{\rm unbound}/M_{\rm tot,0}$, for $\beta^b=0.5$ (solid) and
		$1.5$ (dashed). Time is normalized by the outer pericenter time
		$T_p$ (Table~\ref{tab:actual_orbital_phasing}), so $t/T_p=1$ marks
		the CM passage through the IMBH pericenter. 
	}
	\label{fig:unboundm}
\end{figure}

Figures~\ref{fig:core_mass} and \ref{fig:unboundm} show the bound core
masses, determined using the procedure of Section~\ref{subsec:boundcore},
and the mass bound to neither core, $M_{\rm unbound}$.

For $\beta^b=0.5$, the first internal pericenter produces only modest
changes in the core masses. Sustained mass transfer develops during the
second, strongly perturbed cycle, beginning progressively earlier from
$q=0.2$ to $q=0.7$, following the ordering of the second inner pericenter
passages (Table~\ref{tab:actual_orbital_phasing}). The donor then undergoes
rapid depletion while the accretor grows. For $\beta^b=1.5$, the
$q\geq0.3$ systems enter sustained depletion earlier than their
$\beta^b=0.5$ counterparts and generally show a sharper initial loss of
donor mass. The $q=0.2$, $\beta^b=1.5$ model is the exception. Its donor
has a large bound core, and the accretor grows by a small amount.

Only a small amount of mass is unbound from both cores before the outer
pericenter passage. $M_{\rm unbound}$ then increases but remains at most
a few percent of the initial binary mass, indicating that most stripped
material remains associated with one of the cores or the merger remnant.
The surviving $q=0.2$, $\beta^b=1.5$ model has a much smaller
unbound fraction than its $\beta^b=0.5$ counterpart. 
Since particles can get bound again later, the curves are not monotonic.

\subsubsection{Angular Momentum Redistribution}
\label{subsubsec:angmom}

\begin{figure}[h]
	\epsscale{1.17}
	\plottwo{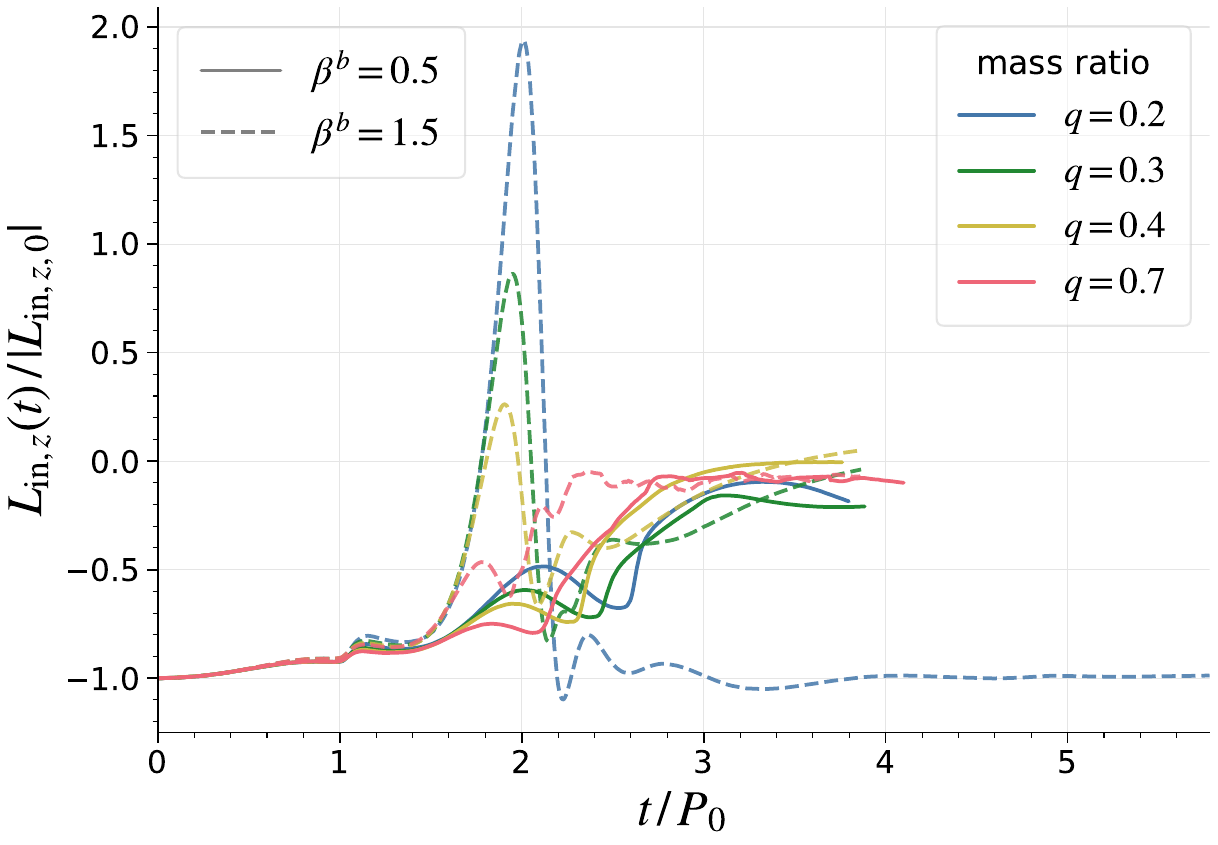}{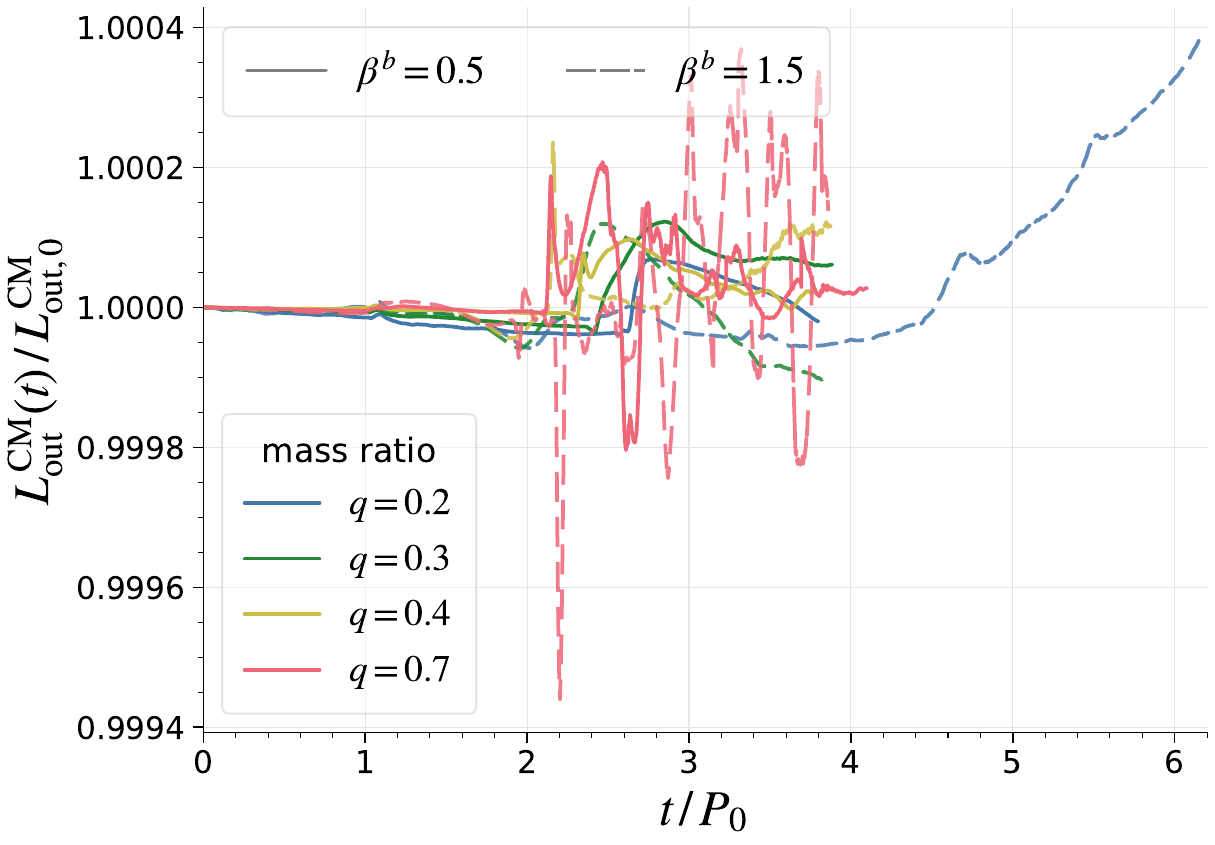}
	\caption{
		Angular momentum evolution for $\beta^b=0.5$ (solid) and $1.5$
		(dashed). \textbf{Left:} Signed $z$-component of the internal orbital
		angular momentum, $L_{{\rm in},z}/|L_{{\rm in},z,0}|$, which spans
		$1.54\times10^{43}$--$3.03\times10^{43}\ {\rm kg \, m^2 \, s^{-1}}$
		across the models. The initially
		retrograde orbit therefore starts at $-1$. The $x$- and $y$-components remain negligible
		($|L_{{\rm in},x}|,|L_{{\rm in},y}|\sim10^{-5}|L_{{\rm in},z,0}|$),
		so the sign of $L_{{\rm in},z}$ directly traces the sense of the
		coplanar internal orbit. \textbf{Right:}
		$L_{\rm out}^{\rm CM}/L_{\rm out,0}^{\rm CM}$ for the binary CM orbit
		about the IMBH. $L_{\rm out,0}^{\rm CM}$ spans
		$2.25\times10^{17}$--$1.53\times10^{18}\ {\rm m^2 \, s^{-1}}$
		across the models. Time is in units of $P_0$.
	}
	\label{fig:Linout}
\end{figure}

The mass-transfer and mass-loss processes described above are accompanied by substantial changes in the angular-
momentum content of the interacting binary. The left panel of Figure~\ref{fig:Linout} shows the signed component
$L_{{\rm in},z}$ (Section~\ref{subsec:energetics_angmom}), normalized by $|L_{{\rm in},z,0}|$, the initially
retrograde binaries therefore start at $-1$. During the first internal
orbit, $L_{{\rm in},z}$ changes only modestly, whereas much stronger
variations develop during the second cycle near the IMBH pericenter.
For the deep encounters, the $q=0.2$, $0.3$, and $0.4$ models cross
$L_{{\rm in},z}=0$ and become temporarily positive, indicating a transient
reversal of the internal orbital sense. Near the zero crossings, the internal orbital angular momentum becomes
very small, corresponding to nearly radial relative motion of the two cores. The effect is strongest
for $q=0.2$, for which
$L_{{\rm in},z}/|L_{{\rm in},z,0}|\simeq+2$ before reversing again and
settling close to its initial retrograde value. The size of this reversal
decreases toward $q=0.4$ and is absent for $q=0.7$, consistent
with the trend of the inner orbital phase at outer pericenter in Table \ref{tab:actual_orbital_phasing}. 
In the merging systems the two-core orbital angular
momentum subsequently ceases to be well defined as the donor core
can no longer be identified.

\begin{figure}[h]
	\epsscale{1.17}
	\plottwo{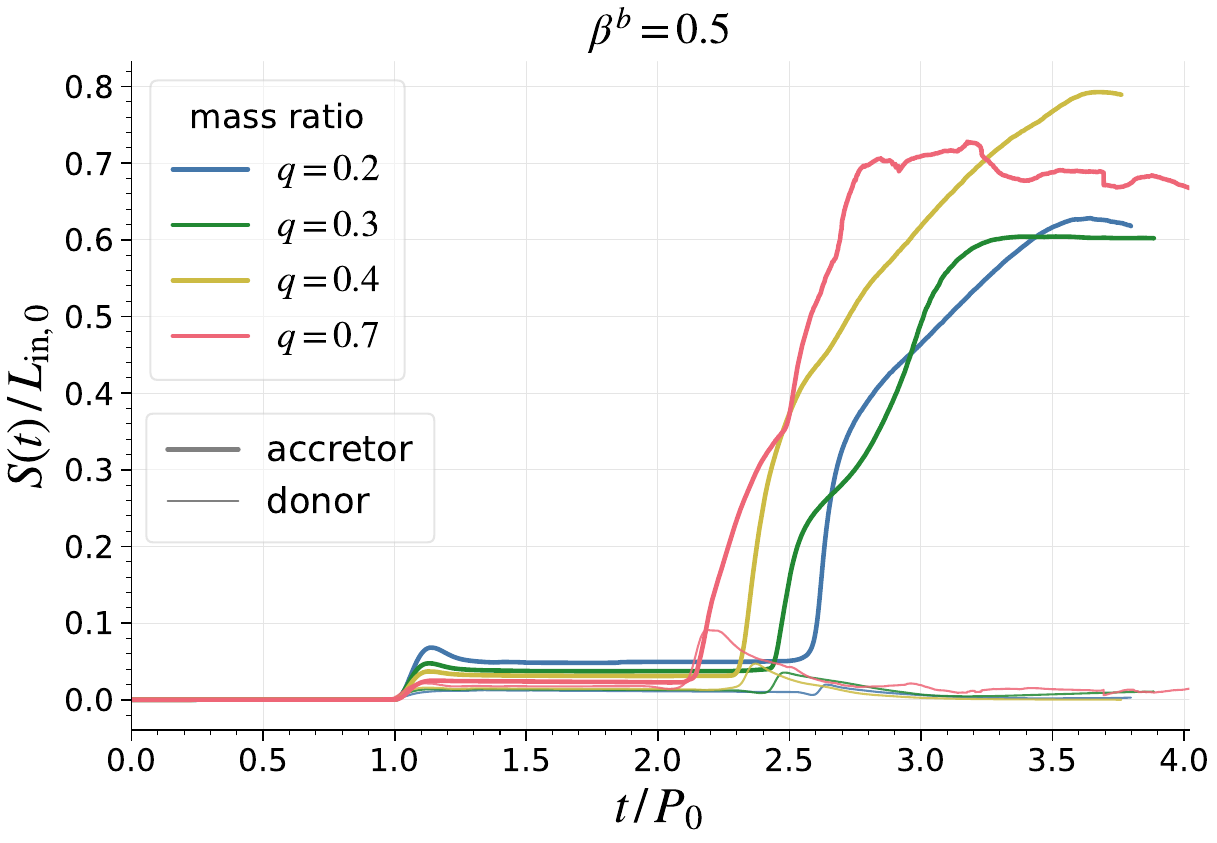}{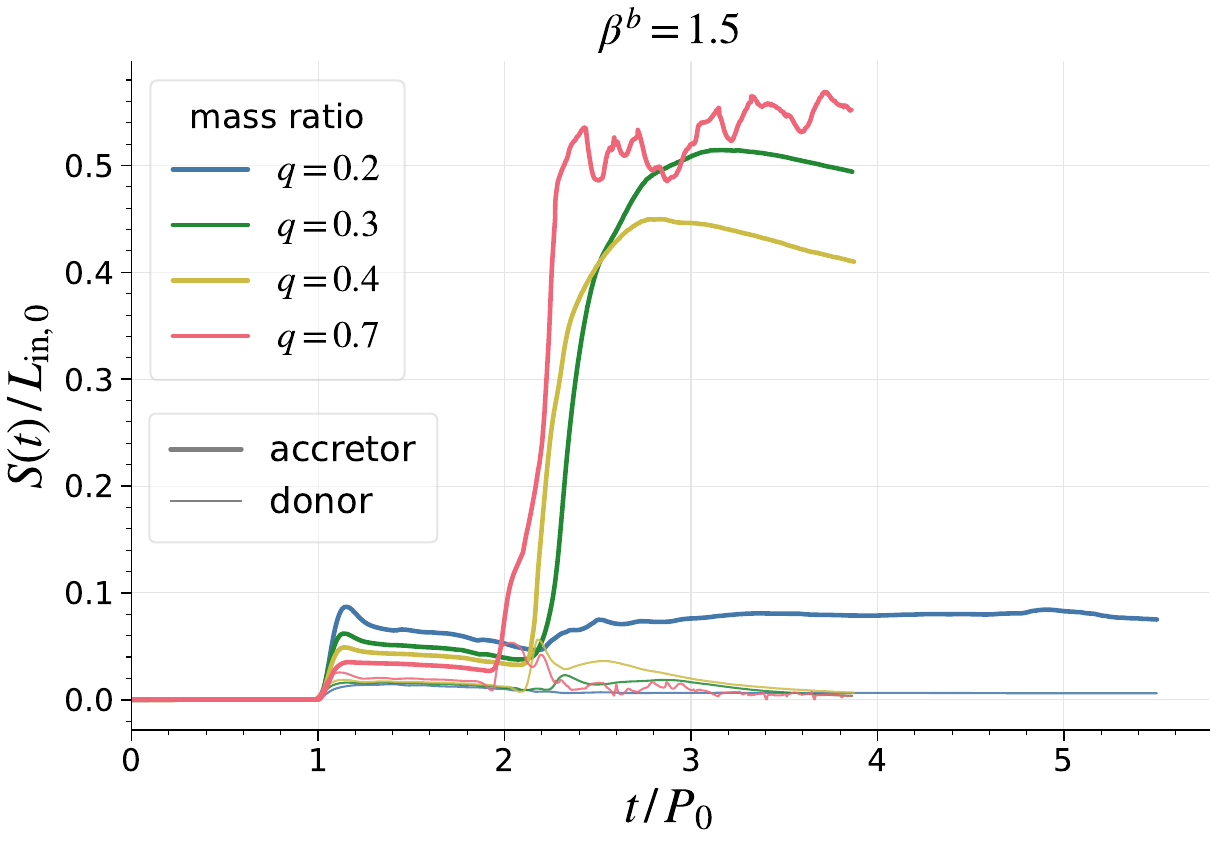}
	\caption{
		Magnitude of the spin angular momentum of material bound to the
		accretor (thick) and donor (thin), normalized by $L_{\rm in,0}$.
		\textbf{Left:} $\beta^b=0.5$. \textbf{Right:} $\beta^b=1.5$.
		Time is in units of $P_0$.
	}
	\label{fig:Spin}
\end{figure}

Both stars are initially non-rotating. The spin angular momentum of the
material bound to each core (Figure~\ref{fig:Spin}) remains small initially
and grows during the strong mass transfer phase. In all models the accretor
gains more spin than the donor, since the transferred material carries angular momentum.
Among the merging configurations,
$S_{\rm acc}/L_{\rm in,0}\simeq0.6$--$0.8$ for $\beta^b=0.5$ and
$\simeq0.4$--$0.6$ for $\beta^b=1.5$. The values are smaller in the deep
encounters possibly because mass transfer lasts for a shorter time before the donor is disrupted. Here $S_{\rm acc}$ includes all material bound
to the accretor, not only the compact WD.

The external orbit's characteristic $L_{\rm out}^{\rm CM}$
(Figure~\ref{fig:Linout}, right) varies by only a few
$\times10^{-4}$ of its initial value. The largest short-timescale
fluctuations occur for $q=0.7$, particularly for $\beta^b=1.5$, while the
surviving $q=0.2$, $\beta^b=1.5$ system shows a slow late-time increase with
small step-like variations. Since the CM is defined from the instantaneous
bound cores, these features can reflect both changes in the external orbit
and changes in the bound core mass distribution. 

\subsubsection{Energetics}
\label{subsubsec:energetics}

\begin{figure}[h]
	\epsscale{1.17}
	\plottwo{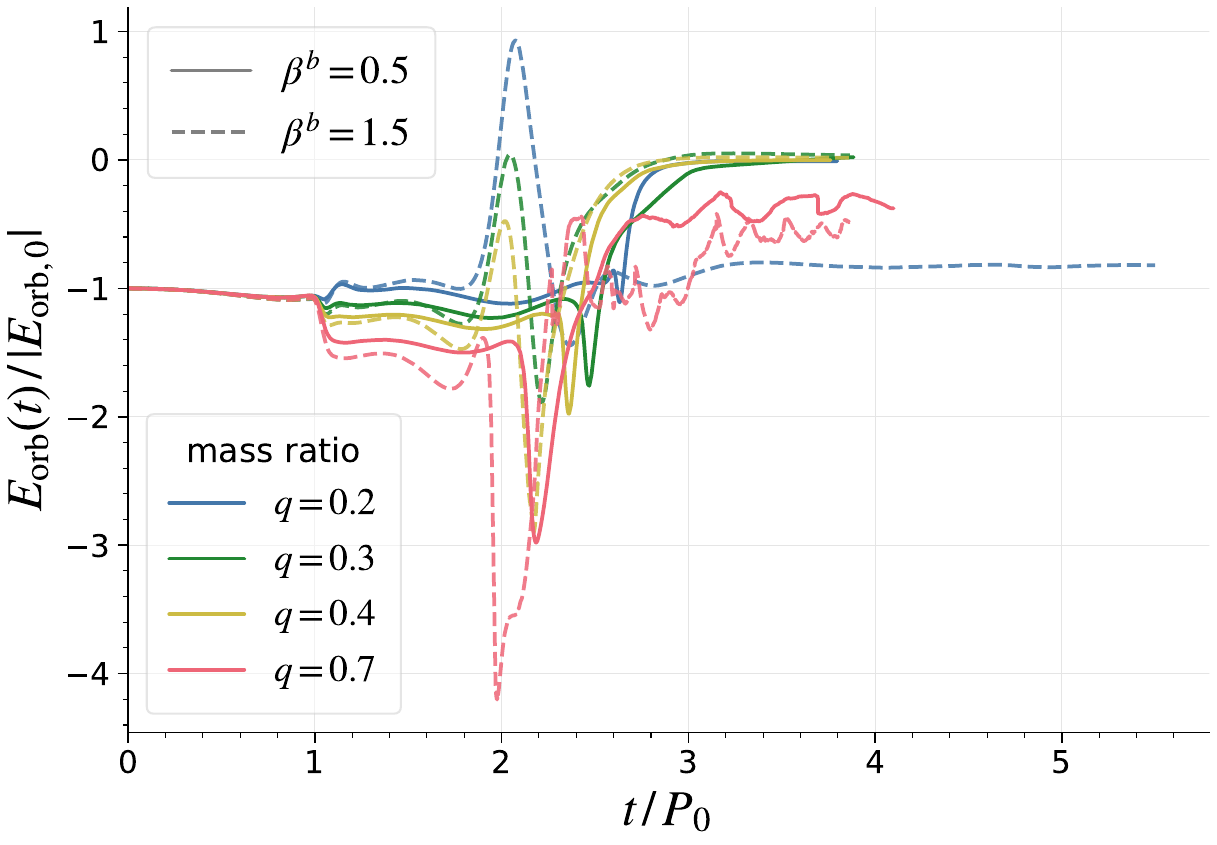}{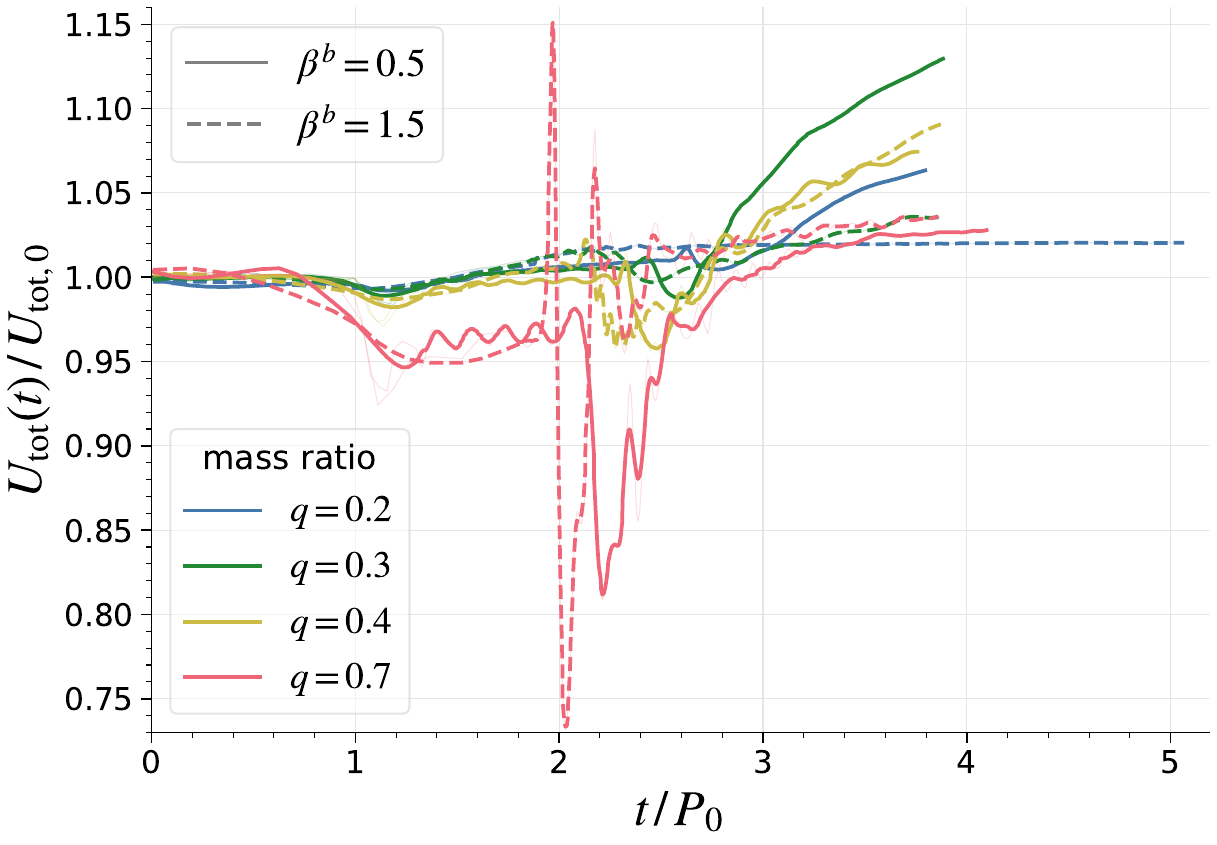}
	\caption{
		Orbital and internal energy for $\beta^b=0.5$ (solid) and $1.5$
		(dashed). \textbf{Left:} Binary orbital energy $E_{\rm orb}$,
		normalized by $|E_{\rm orb,0}|=3.88\times10^{40}$--$3.04\times10^{41}$\,J.
		\textbf{Right:} Internal energy of material bound to the two cores,
		$U_{\rm tot}$, normalized by
		$U_{\rm tot,0}=(3.30$--$4.49)\times10^{42}$\,J.
		Time is in units of $P_0$.
	}
	\label{fig:energy}
\end{figure}

To characterize the energetic response of the binary during the IMBH
encounter, Figure~\ref{fig:energy} shows the orbital energy of the two bound cores,
$E_{\rm orb}$, and the internal energy of material bound to them,
$U_{\rm tot}$ (Section~\ref{subsec:energetics_angmom}).

The largest changes in $E_{\rm orb}$ occur during the strongly perturbed
second cycle. In the merging $q=0.3$--$0.7$ models, $E_{\rm orb}$ becomes
most negative near the second internal pericenter, with deeper minima at
larger $q$. In the $\beta^b=1.5$ encounters, $E_{\rm orb}$ briefly becomes positive
near the outer pericenter, with $E_{\rm orb}>0$ for $q=0.2$
and $0.3$. After the donor core disappears,
$E_{\rm orb}$ no longer represents a two-body orbital energy.

The internal energy changes by only a few percent before the second cycle
and varies much more during mass transfer and disruption. The
largest short term change occurs for $q=0.7$, $\beta^b=1.5$.
Since $U_{\rm tot}$ includes only material assigned to the two bound cores,
these variations come partly from heating and partly from particles joining/leaving the core. In the merging models,
$U_{\rm tot}$ remains somewhat above its initial value due to compression
and heating.

The surviving $q=0.2$, $\beta^b=1.5$ model behaves differently.
After becoming positive, $E_{\rm orb}$ returns to
$\simeq-0.82\,|E_{\rm orb,0}|$, while $U_{\rm tot}$ remains within
$\simeq2\%$ of its initial value. Together with the settled core masses
and $|L_{{\rm in},z}|\simeq|L_{{\rm in},z,0}|$, this gives
$a_{\rm in}\simeq1.03\,a_{\rm in,0}$ and $e_{\rm in}\simeq0.37$,
consistent with section \ref{subsubsec:orbsep}.

\section{Discussion \& Conclusions}
\label{discussion}

Our simulations show that an IMBH encounter can strongly accelerate the
evolution of an eccentric, mass-transferring WD binary. In isolation,
repeated pericenter passages drive episodic mass transfer and gradual
orbital evolution, and only the $q=0.7$ binary merges within the
simulated interval, at $t/P_0\simeq15$. In the presence of the IMBH, all
but one configuration merge during or shortly after the second internal
pericenter passage. The exception is the $q=0.2$, $\beta^b=1.5$
configuration, which retains two bound cores throughout the simulated
interval and settles onto an orbit with nearly the same semimajor axis
but a lower eccentricity, $e_{\rm in}\simeq0.4$.

The outcome also appears sensitive to the inner-orbital phase at which the
binary reaches outer pericenter. At fixed $\beta^b$, the outer-pericenter
time in units of $P_0$ is nearly independent of $q$, whereas the second
inner pericenter shifts systematically with mass ratio. The strongest
tidal perturbation therefore acts at different phases of the inner orbit.
This phase dependence is consistent with the different outcomes of the
$q=0.2$ binaries at $\beta^b=0.5$ and $1.5$, although the encounter
strength also differs. In the deep encounters with $q=0.2$, $0.3$, and
$0.4$, $L_{{\rm in},z}$ temporarily changes sign, corresponding to a
transient reversal of the internal orbital sense. The effect is strongest
for the surviving $q=0.2$ model, which subsequently returns to a
retrograde orbit with $|L_{{\rm in},z}|$ close to its initial value.

Mass transfer also redistributes angular momentum within the binary.
The accretor gains more spin than the donor, consistent
with transferred material carrying angular momentum into the accretor and
its surrounding bound material. In contrast, the orbital angular momentum
of the binary CM about the IMBH changes by only a few times
$10^{-4}$. The encounter thus changes the inner binary strongly, without 
much change in its orbit around the IMBH. 

The post-processing analysis of Appendix~\ref{nuclear_ignition} identifies
He-origin material that are hot and dense enough for rapid helium burning.
No such material is found for $q\leq0.3$, while the candidate fraction is
$\sim2\%$ for $q=0.4$ and $\sim15$--$17\%$ for $q=0.7$, where it is
concentrated around the CO core. No CO-origin material satisfies the
carbon-burning criterion. Whether this leads to helium detonation can be
tested with simulations that include a nuclear network. 

The present survey samples a restricted region of parameter space, with
four mass ratios, two encounter strengths, a fixed inner eccentricity, and
an initial configuration chosen to undergo Roche-lobe overflow near inner
pericenter. A broader survey in mass ratio, encounter strength, eccentricity, and the
relative phase between the inner and outer orbits will be needed to
determine how generally the merger and two-core outcomes found here
occur. Such a phase study can retain the present initialization of the
inner binary at apocenter while varying the time at which the IMBH
pericenter is encountered.

\begin{center}{\bf Acknowledgments}\end{center}

We acknowledge the support and resources provided by PARAM Sanganak under the National Supercomputing Mission, Government of
India, at the Indian Institute of Technology Kanpur. We thank Sabyasachi Chakraborty for generously allowing us to use his computer cluster.  

\begin{center}{\bf Data Availability Statement}\end{center}

The data underlying this article will be shared upon reasonable request of the corresponding author.

\appendix

\section{Thermodynamic State and Possible Dynamical Burning}
\label{nuclear_ignition}
Mass transfer and merger strongly heat the disrupted material, particularly where donor debris falls onto the primary. Since
nuclear reactions are not coupled to the hydrodynamics, we check separately whether this material reaches conditions in which
nuclear heating could become dynamically important, by comparing local nuclear heating and dynamical timescales for each SPH
particle. 

For He-origin material ($X_{\rm He}=1$), the triple $\alpha$ heating timescale is
\begin{equation}
	\tau_{3\alpha}
	=
	\frac{c_P T}{\dot{\epsilon}_{3\alpha}},
	\label{eq:tau3alpha}
\end{equation}
where the screened triple $\alpha$ energy-generation rate is \citep{Dan2014}
\begin{equation}
	\dot{\epsilon}_{3\alpha}
	=
	5.09\times10^{11}
	f_{3\alpha}\rho^2 X_{\rm He}^{3}T_8^{-3}
	\exp\left(-\frac{44.027}{T_8}\right)
	\ \mathrm{erg \, g^{-1} \, s^{-1}},
    \label{eq:eps3alpha}
\end{equation}
with
\begin{equation}
	f_{3\alpha}
	=
	\exp\left(
	2.76\times10^{-3}\rho^{1/2}T_8^{-3/2}
	\right),
	\qquad
	T_8=\frac{T}{10^8\,\mathrm{K}} .
	\label{eq:screen3alpha}
\end{equation}
For this post-processing calculation, we evaluate the constant-pressure
specific heat $c_P$ using contributions from zero-temperature degenerate
electrons, ideal ions, and radiation \citep{Benz1989}. Specifically,
\begin{equation}
	c_P
	=
	c_V
	+
	\frac{T}{\rho^2}
	\frac{
		\left[(\partial P/\partial T)_\rho\right]^2
	}{
		(\partial P/\partial\rho)_T
	},
	\label{eq:cp}
\end{equation}
where $c_V$ is the specific heat at constant volume. Finite-temperature
electron corrections are not included in this post-processing calculation. For CO-origin material ($X_{\rm C}=0.5$), we use the analytic carbon-burning rate and self-acceleration correction of \citet{Dan2014}.

Following \citet{Dan2014}, particles with $\tau_{3\alpha}<\tau_{\rm dyn}=(G\rho)^{-1/2}$ are classified as candidates for
dynamical He burning. To test the sensitivity to this choice, we also use the shorter timescale $(24\pi G\rho)^{-1/2}$ adopted in
related WD-merger studies \citep{Sato2015}. Because individual SPH temperatures fluctuate on the particle scale, the classification uses a kernel-smoothed temperature, following the SPH-smoothed
temperature of \citet{Dan2012} but evaluated with the M6 kernel and normalized by
the kernel sum,
\begin{equation}
	T_i^{\rm M6}
	=
	\frac{
		\displaystyle\sum_j (m_j/\rho_j)\,T_j W_{ij}
	}{
		\displaystyle\sum_j (m_j/\rho_j)\,W_{ij}
	},
	\label{eq:tm6}
\end{equation}
where the sum runs over neighbors within the M6 kernel support. The raw particle temperatures serve as a consistency check. We
evaluate all models at $t/P_0\simeq3.5$, after the strongest mass transfer phase in the merging systems. This epoch was
fixed from the hydrodynamic evolution, independently of the temperatures. Candidate material is still present 
in later outputs, but the simulations do not extend far enough to find its
maximum, so the quoted masses are instantaneous, and not peak values. For the surviving $q=0.2$, $\beta^b=1.5$ model, this
snapshot is a late-time state rather than a merger remnant.

\begin{figure}[!htbp]
	\centering

	\gridline{
		\fig{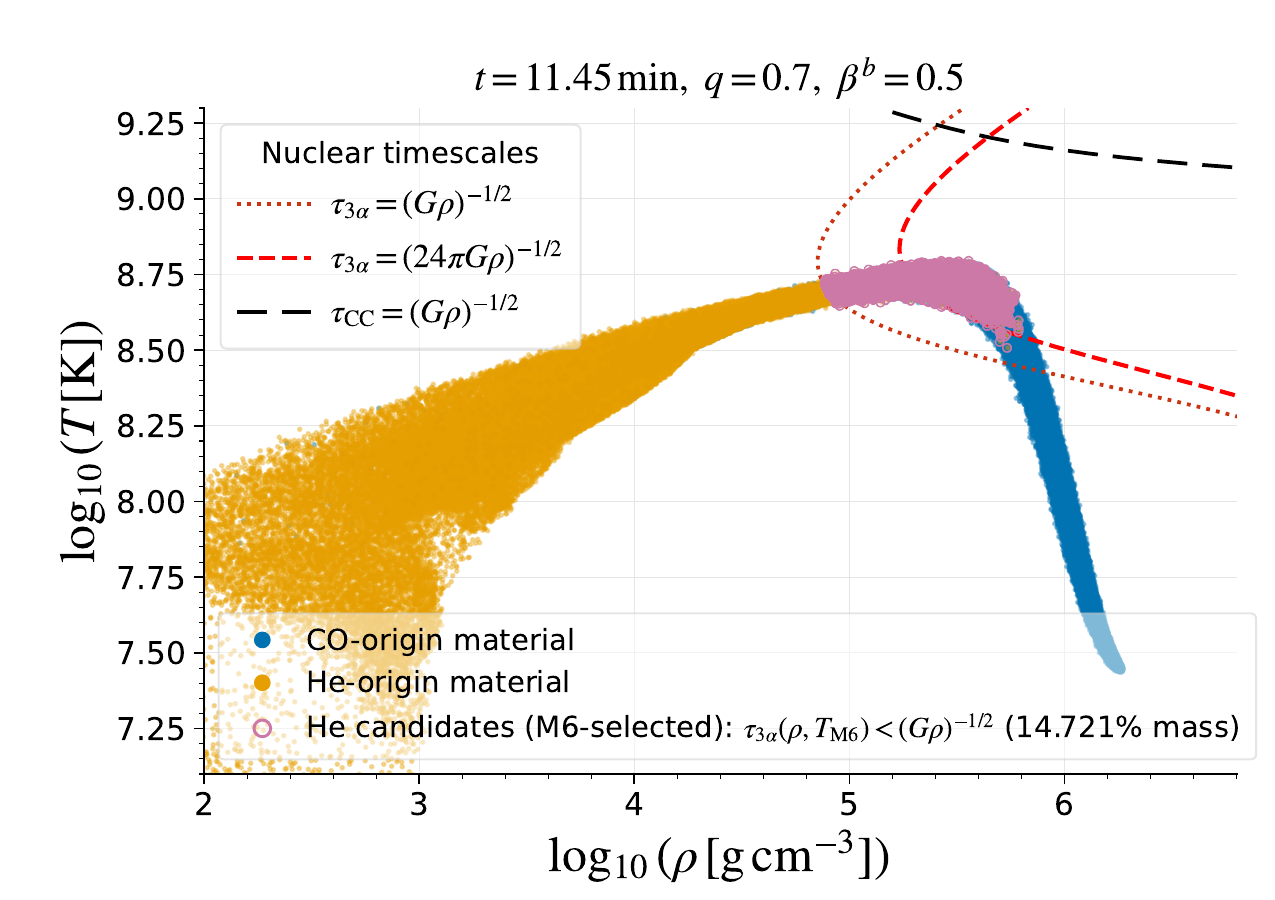}{0.51\textwidth}{}
		\fig{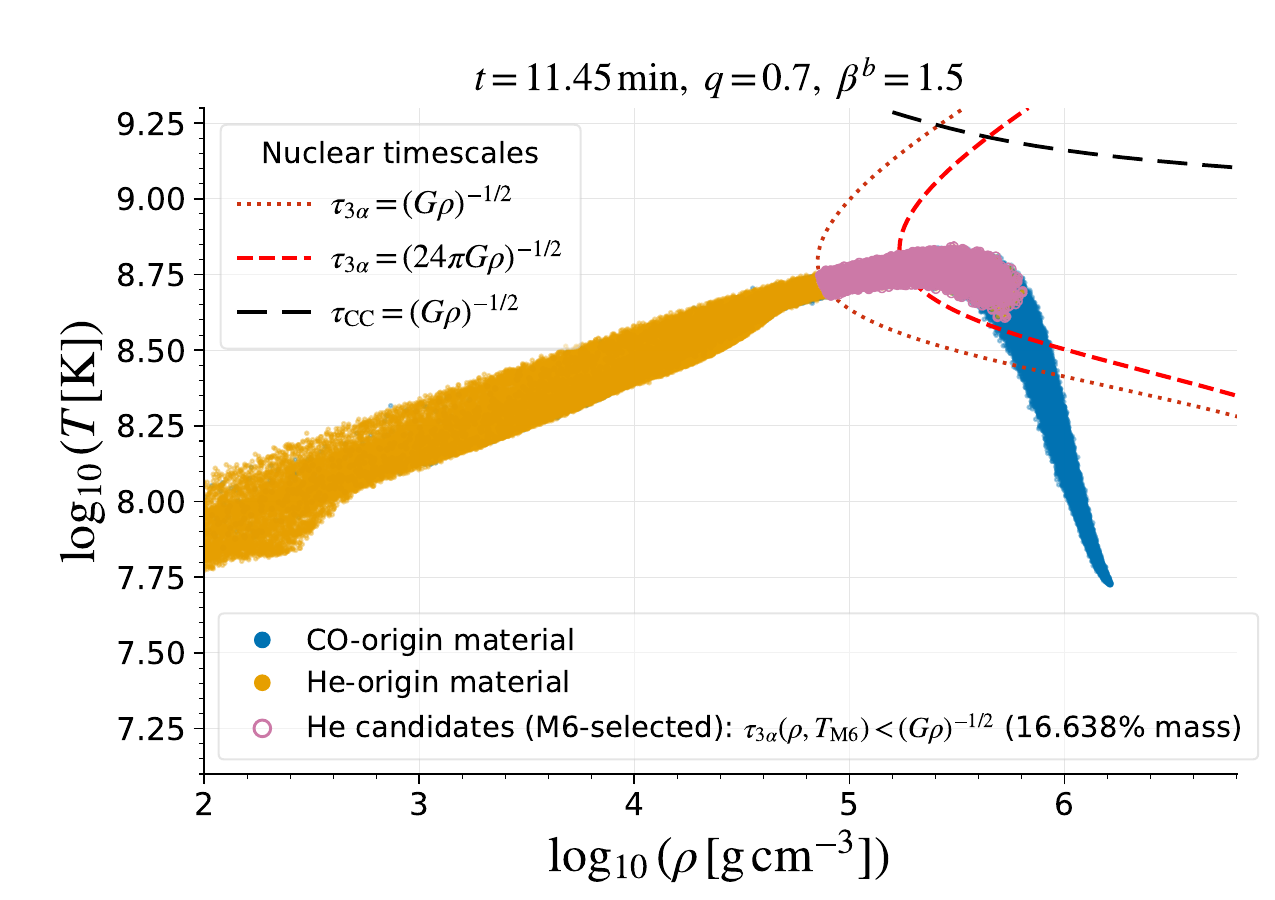}{0.51\textwidth}{}
	}
	\vspace{-1.5cm}
	\gridline{
		\fig{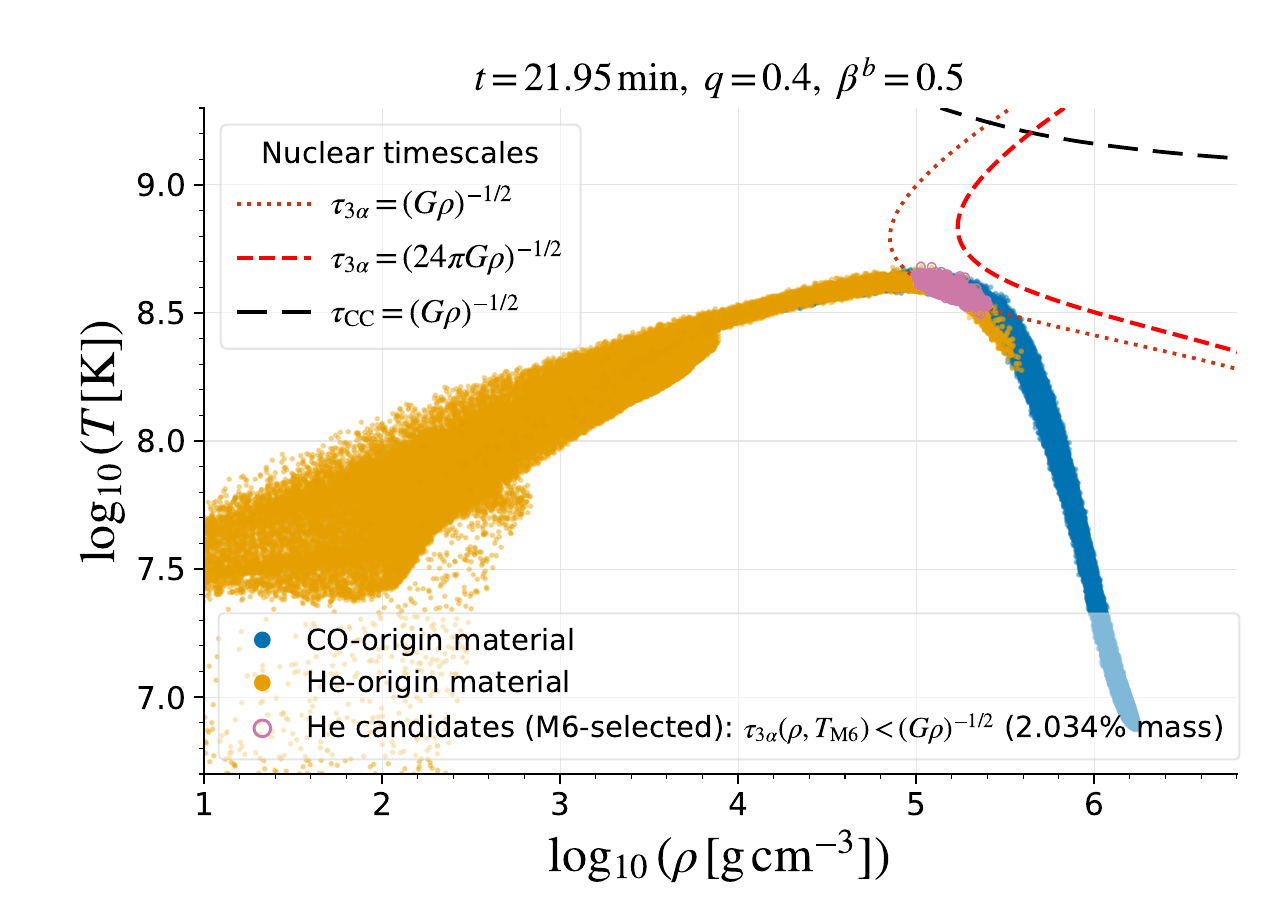}{0.51\textwidth}{}
		\fig{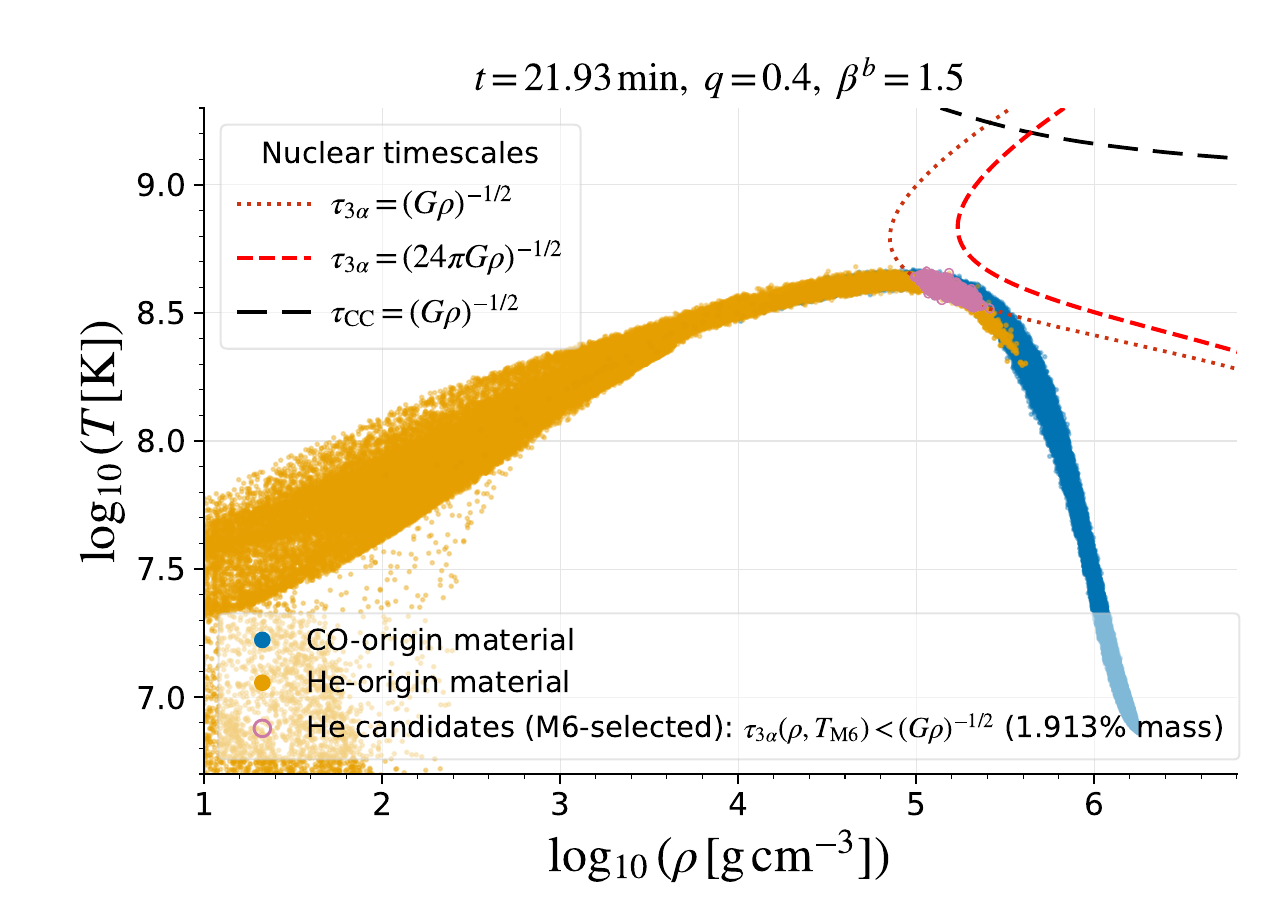}{0.51\textwidth}{}
	}
	\vspace{-0.6cm}
	\caption{ Density--temperature distributions at $t/P_0\simeq3.5$ for $q=0.7$ (top) and $q=0.4$ (bottom), with $\beta^b=0.5$ (left)
	and $1.5$ (right). Blue and orange points show CO- and He-origin material, plotted with raw temperatures. Open purple circles mark
	He-origin candidates, selected with $T_i^{\rm M6}$ and $\tau_{3\alpha}<(G\rho)^{-1/2}$. Red dotted and dashed curves show
	$\tau_{3\alpha}=(G\rho)^{-1/2}$ and $(24\pi G\rho)^{-1/2}$, and the black dashed curve shows $\tau_{\rm CC}=(G\rho)^{-1/2}$.
	}
	\label{fig:rho_T}
\end{figure}
Part of the heated He-origin material lies where $\tau_{3\alpha}$ is shorter than the dynamical time (Figure~\ref{fig:rho_T}),
whereas no CO-origin particle meets the carbon-burning criterion in any model. The candidate fraction depends much more on mass
ratio than on encounter strength. It is $\simeq15$--$17\%$ of the He-origin mass for $q=0.7$, $\simeq2\%$ for $q=0.4$, and zero
for $q\leq0.3$. For $q=0.7$, $\beta^b=1.5$, the fraction is $16.64\%$ with either raw or M6-smoothed temperatures and falls to
$6.45\%$ and $6.43\%$, respectively, with the $(24\pi G\rho)^{-1/2}$ criterion. The result is therefore insensitive to temperature
smoothing but depends on the choice of dynamical timescale. The
quoted candidate fractions should be the upper limits as finite temperature correction of electron is not included.

\begin{figure}[!htbp]
	\epsscale{0.6}
	\plotone{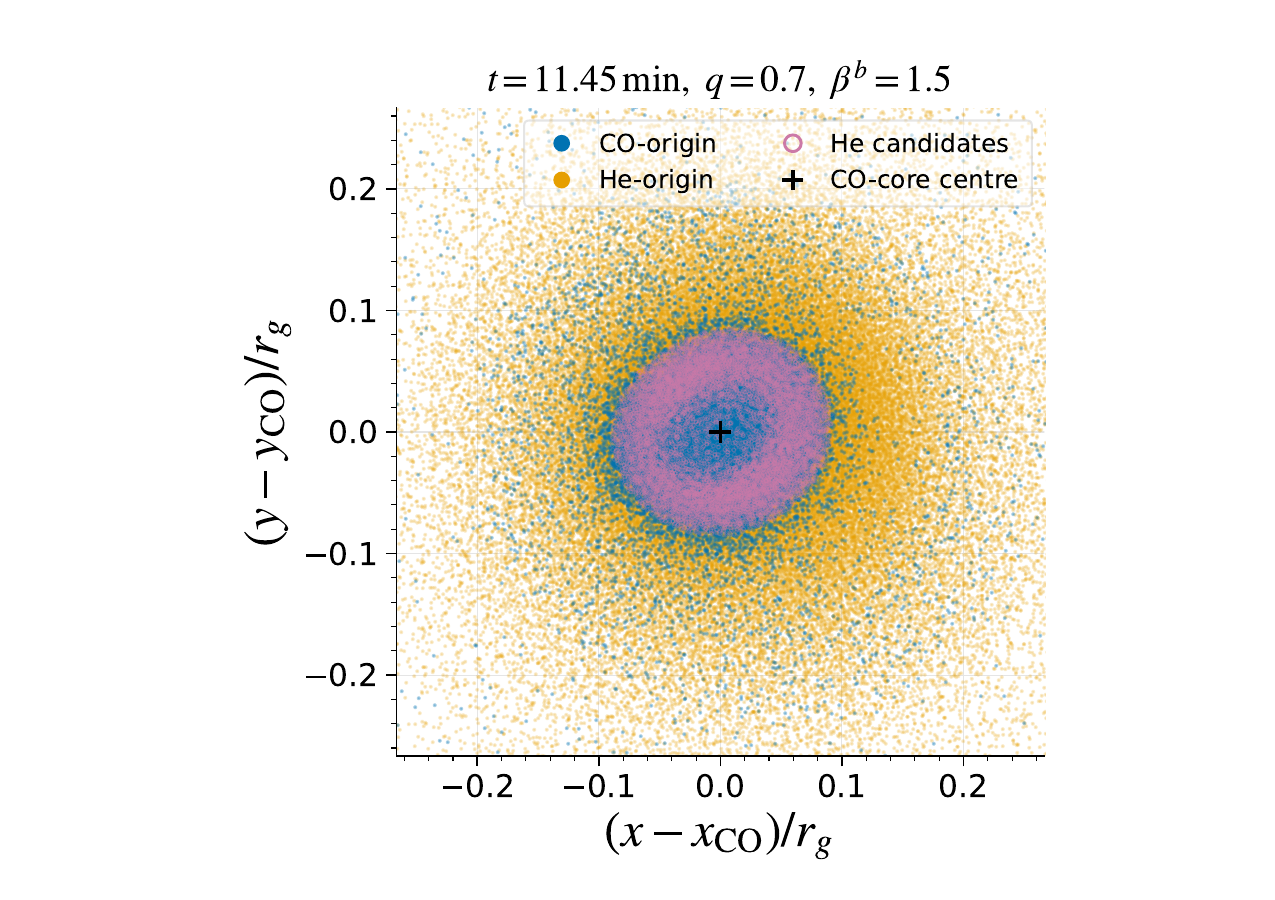}
	\caption{ Material around the CO core in the $q=0.7$, $\beta^b=1.5$ model at $t/P_0\simeq3.5$, in units of $r_g$ relative to the
	core center (black plus). Blue and orange points show CO- and He-origin particles, and open purple circles mark He-origin
	candidates with $\tau_{3\alpha}(\rho_i,T_i^{\rm M6})<(G\rho_i)^{-1/2}$. The identified CO core has a radius of about $0.087\,r_g$.
	}
	\label{fig:burning_spatial}
\end{figure}
In this model, the $5.82\times10^{-2}\,M_\odot$ of candidate material is concentrated around the CO core and 
not scattered through the debris (Figure~\ref{fig:burning_spatial}). To test whether local expansion could
quench the burning, we estimate an expansion timescale from the SPH velocity divergence,
\begin{equation}
	\tau_{\rm exp}
	=
	(\boldsymbol{\nabla}\cdot\boldsymbol{v})^{-1},
	\qquad
	\boldsymbol{\nabla}\cdot\boldsymbol{v}>0 ,
	\label{eq:tauexp}
\end{equation}
applied only to expanding material. Of the candidate mass, $47.25\%$ is expanding and $52.75\%$ is compressing. Within the
expanding part, $99.50\%$ has $\tau_{3\alpha}<\tau_{\rm exp}$, with a median $\tau_{3\alpha}/\tau_{\rm exp}=0.042$ and a 5--95
percentile range of $0.0018$--$0.484$. Expansion is therefore too slow to quench burning in nearly all of the expanding candidate
material.

These results show that rapid helium burning is possible. Reactions are not coupled to hydrodynamics and the composition
is not evolved, so we cannot say if the burning would spread beyond the candidate material. 
Detonation also requires a sufficiently large, resolved hotspot whose critical size depends on density and
temperature \citep{Holcomb2013}, and burning criteria in SPH simulations of WD--IMBH encounters can be
resolution dependent \citep{Tanikawa2017}. Higher-resolution simulations with a nuclear network are needed to determine whether a
helium detonation develops and to quantify its nucleosynthetic yields.

\section{Numerical Resolution and Convergence}
\label{app:resolution}

\begin{figure}[!htbp]
	\epsscale{1.17}
	\plottwo{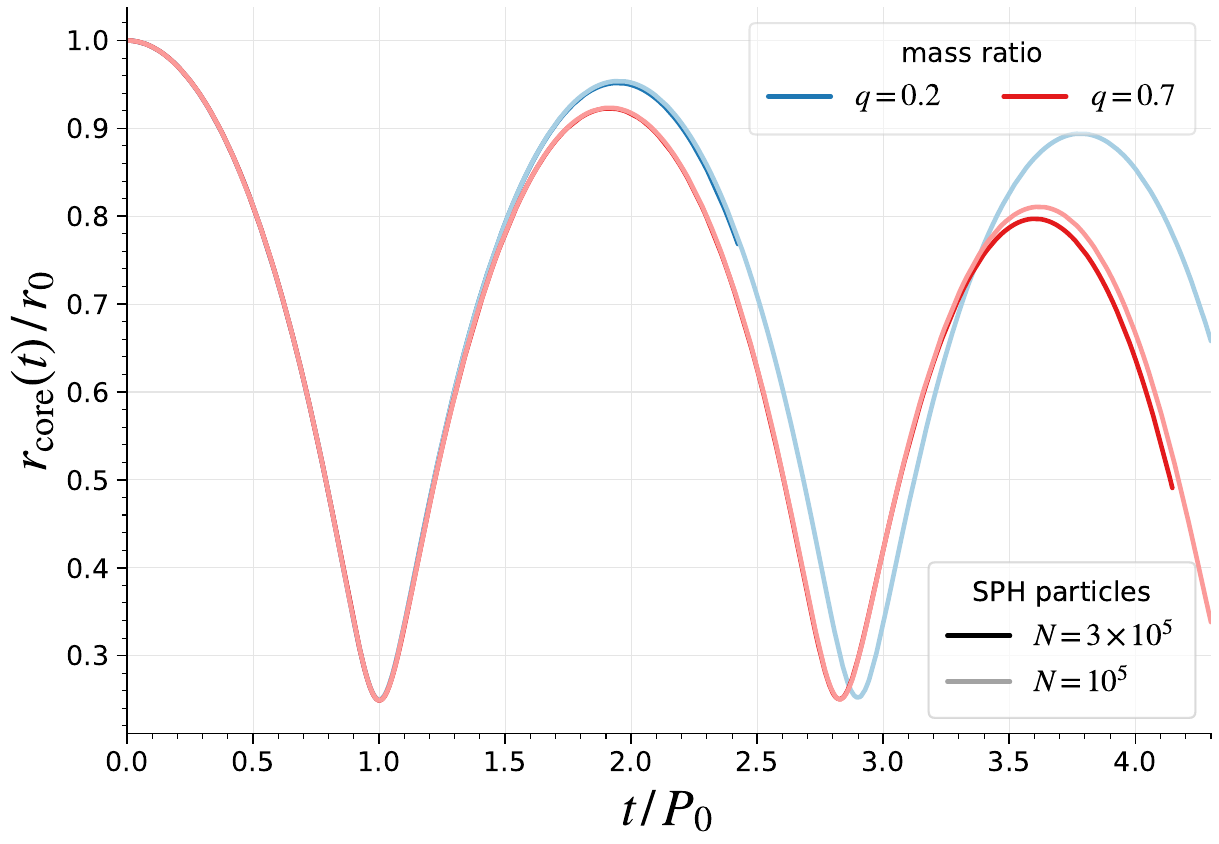}{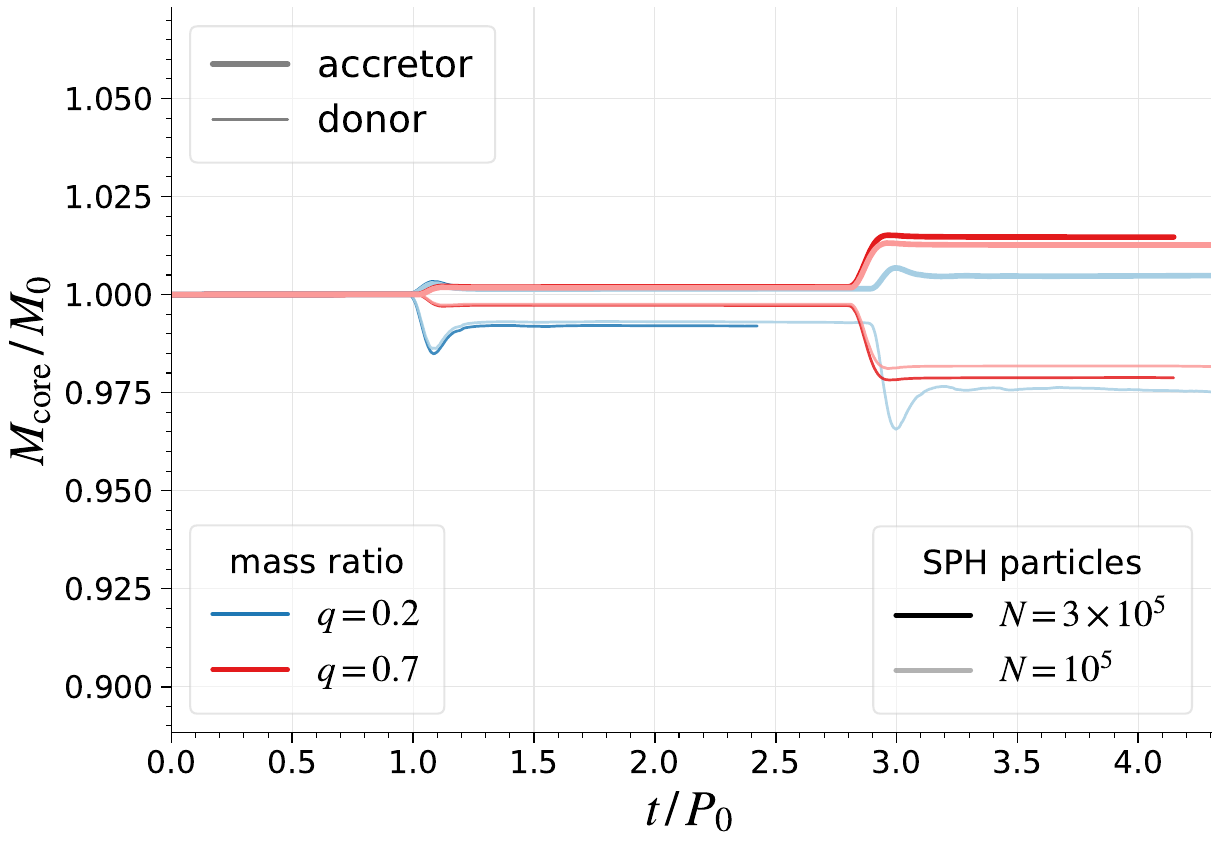}
	\caption{ Isolated $q=0.2$ (blue) and $q=0.7$ (red) binaries. \textbf{Left:} bound core separation $r_{\rm core}/r_0$.
	\textbf{Right:} bound core masses $M_{\rm core}/M_0$ for the accretor (thick) and donor (thin).
	}
	\label{fig:convergence1}
\end{figure}

\begin{figure}[!htbp]
	\epsscale{0.6}
	\plotone{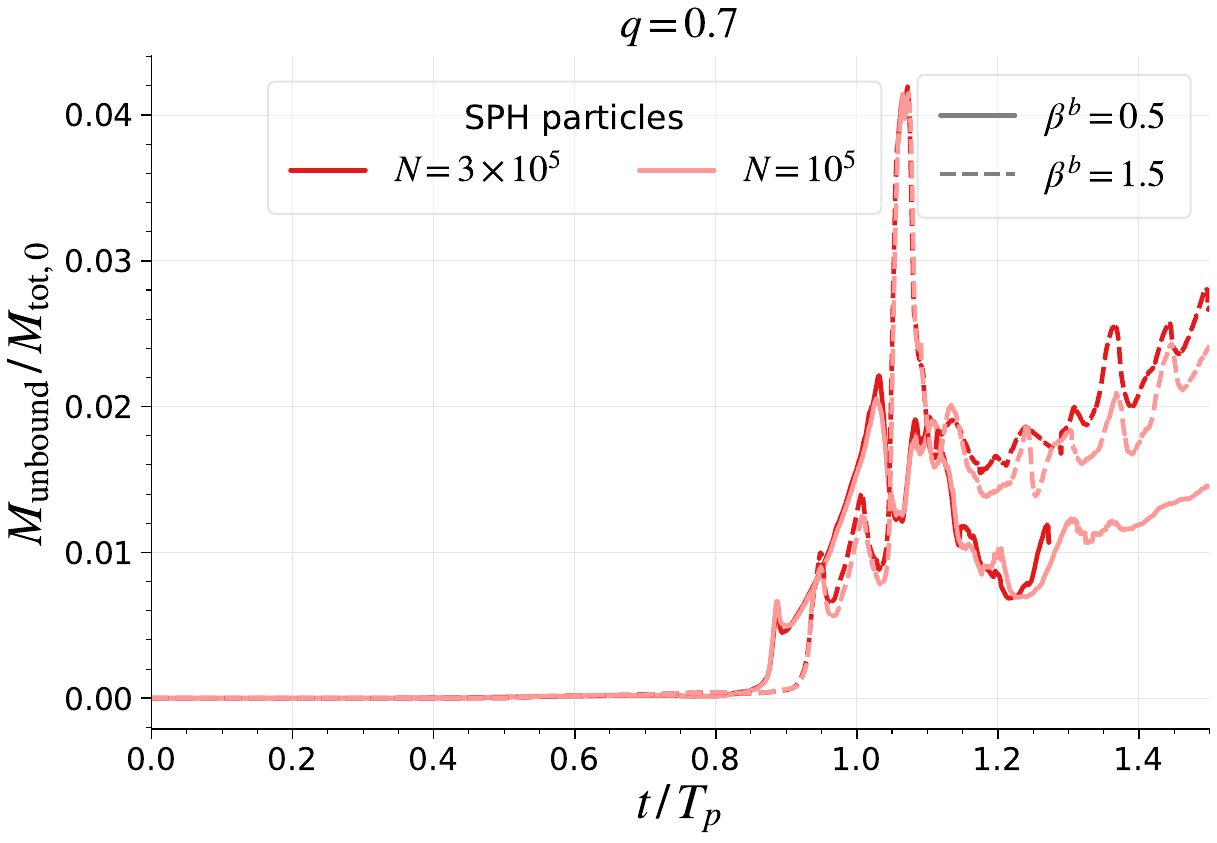}
	\caption{Mass unbound from both cores, $M_{\rm unbound}/M_{\rm tot,0}$, for the $q=0.7$ encounters, with time in units of $T_p$.
	The higher-resolution $\beta^b=0.5$ run ends at $t/T_p\simeq1.27$.}
	\label{fig:convergence2}
\end{figure}

\begin{figure}[!htbp]
	\epsscale{1.17}
	\plottwo{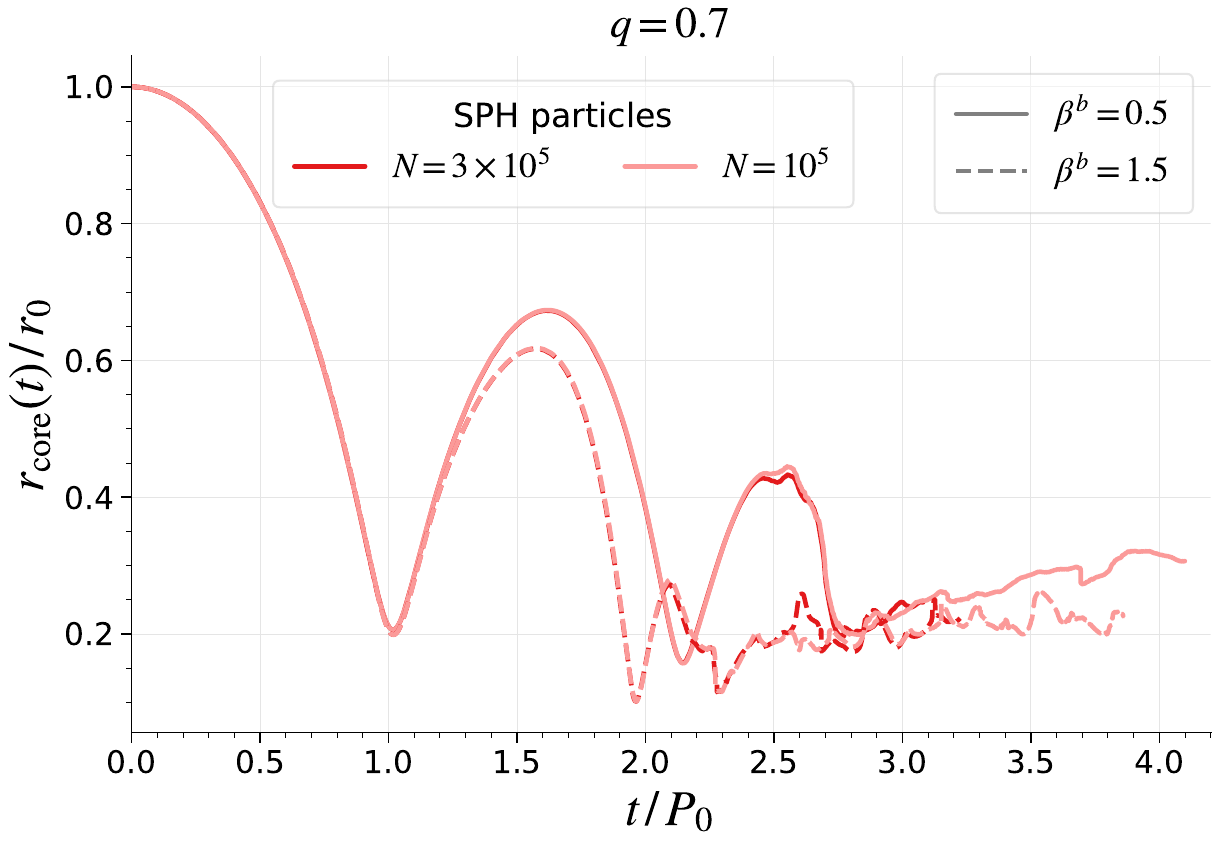}{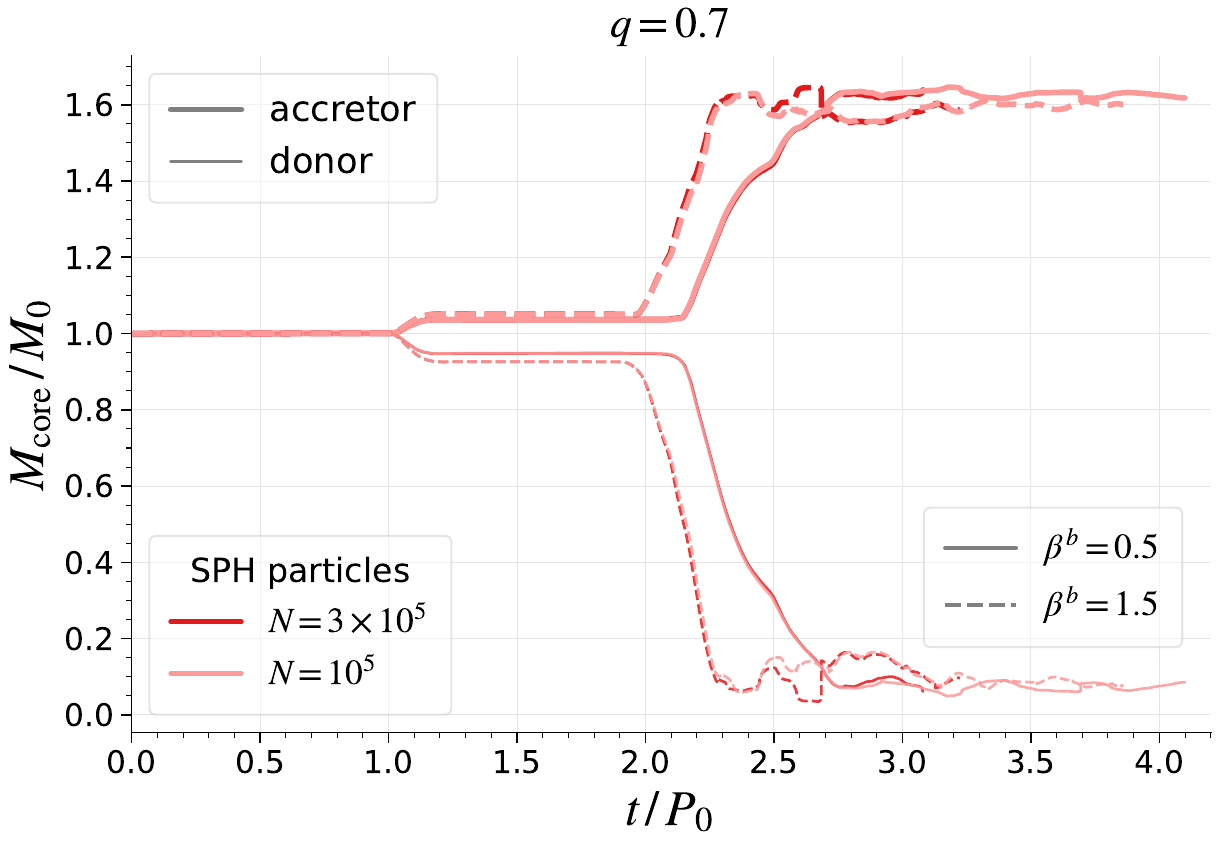}
	\caption{ The $q=0.7$ encounters. \textbf{Left:} bound core separation $r_{\rm core}/r_0$. \textbf{Right:} bound core masses for
	the accretor (thick) and donor (thin).
	}
	\label{fig:convergence3}
\end{figure}

\begin{figure}[!htbp]
	\epsscale{1.17}
	\plottwo{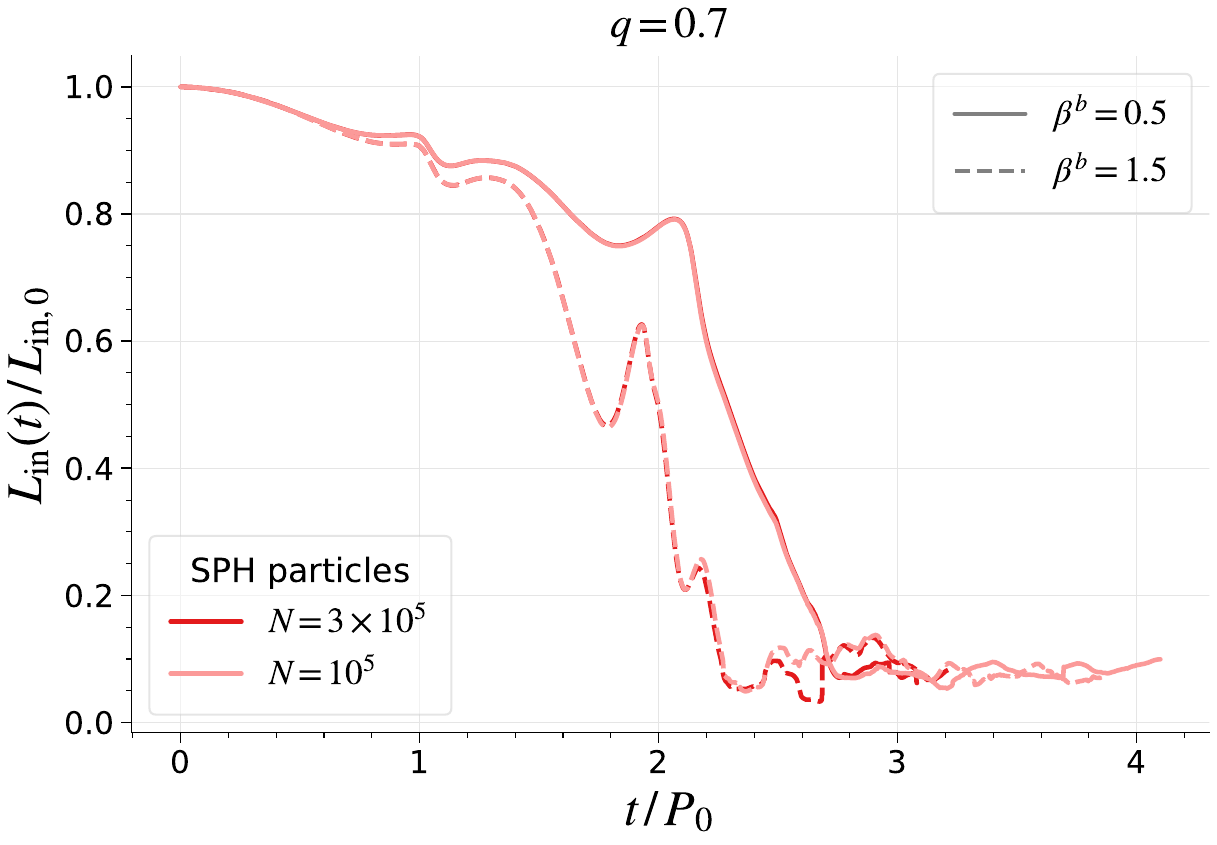}{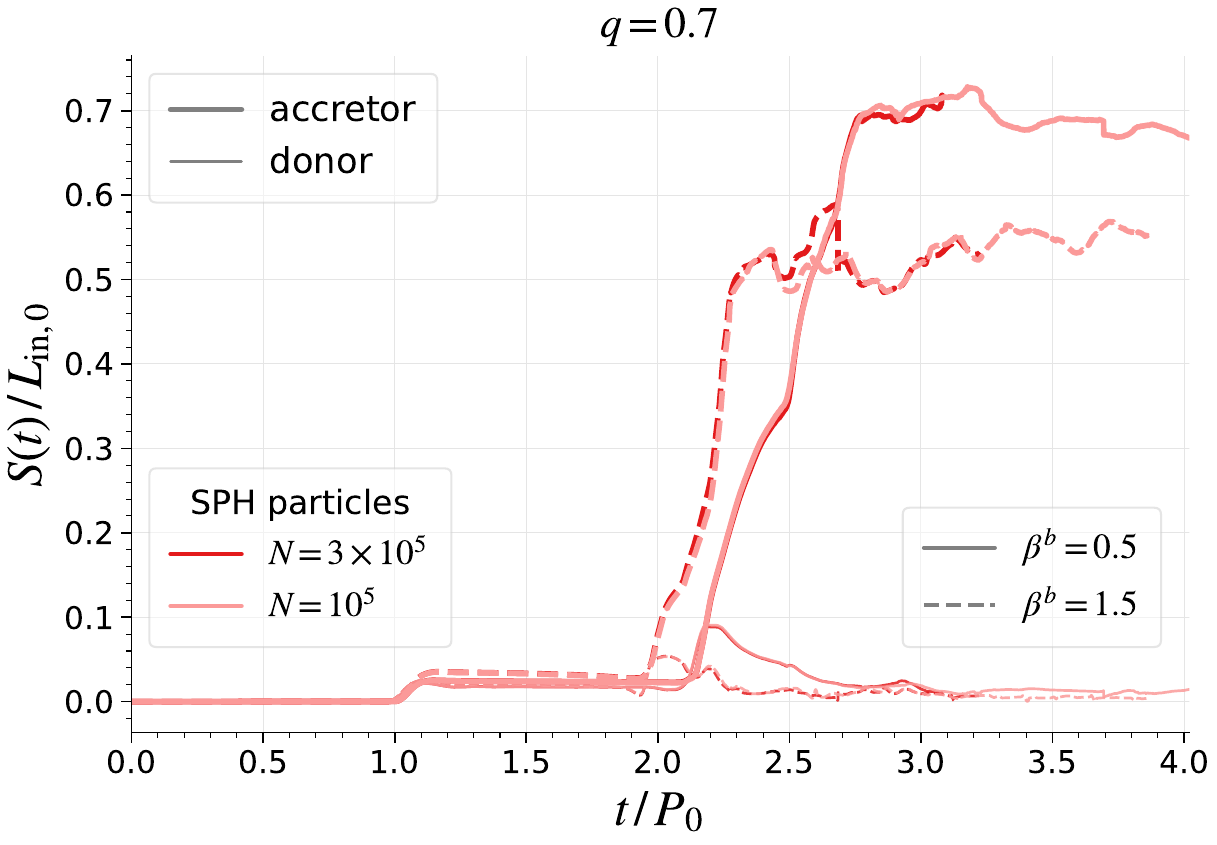}
	\caption{ The $q=0.7$ encounters. \textbf{Left:} $L_{\rm in}/L_{\rm in,0}$. \textbf{Right:} spin of the material bound to the
	accretor (thick) and donor (thin), $S/L_{\rm in,0}$.
	}
	\label{fig:convergence4}
\end{figure}

\begin{figure}[!htbp]
	\epsscale{1.17}
	\plottwo{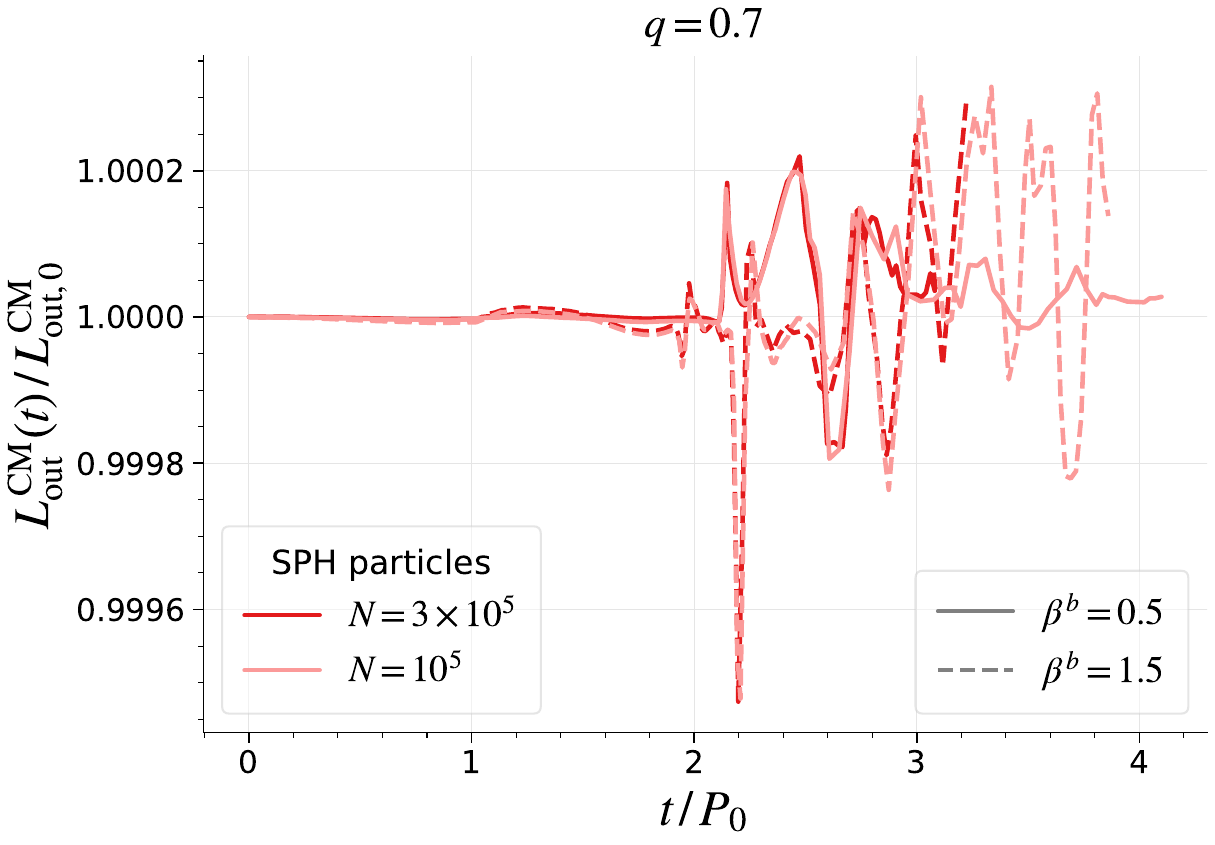}{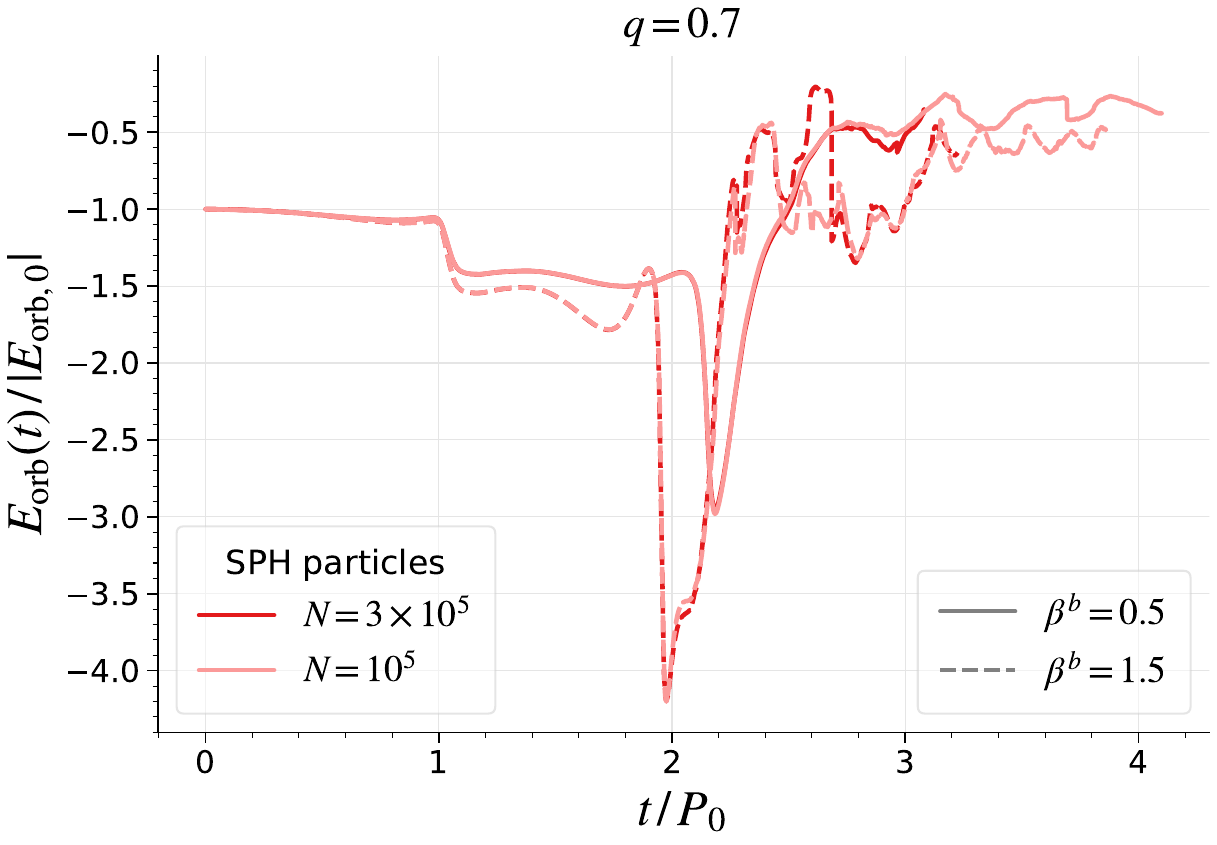}
	\caption{ The $q=0.7$ encounters. \textbf{Left:} $L_{\rm out}^{\rm CM}/L_{\rm out,0}^{\rm CM}$. \textbf{Right:} $E_{\rm
	orb}/|E_{\rm orb,0}|$.
	}
	\label{fig:convergence5}
\end{figure}

\begin{figure}[!htbp]
	\epsscale{1.17}
	\plottwo{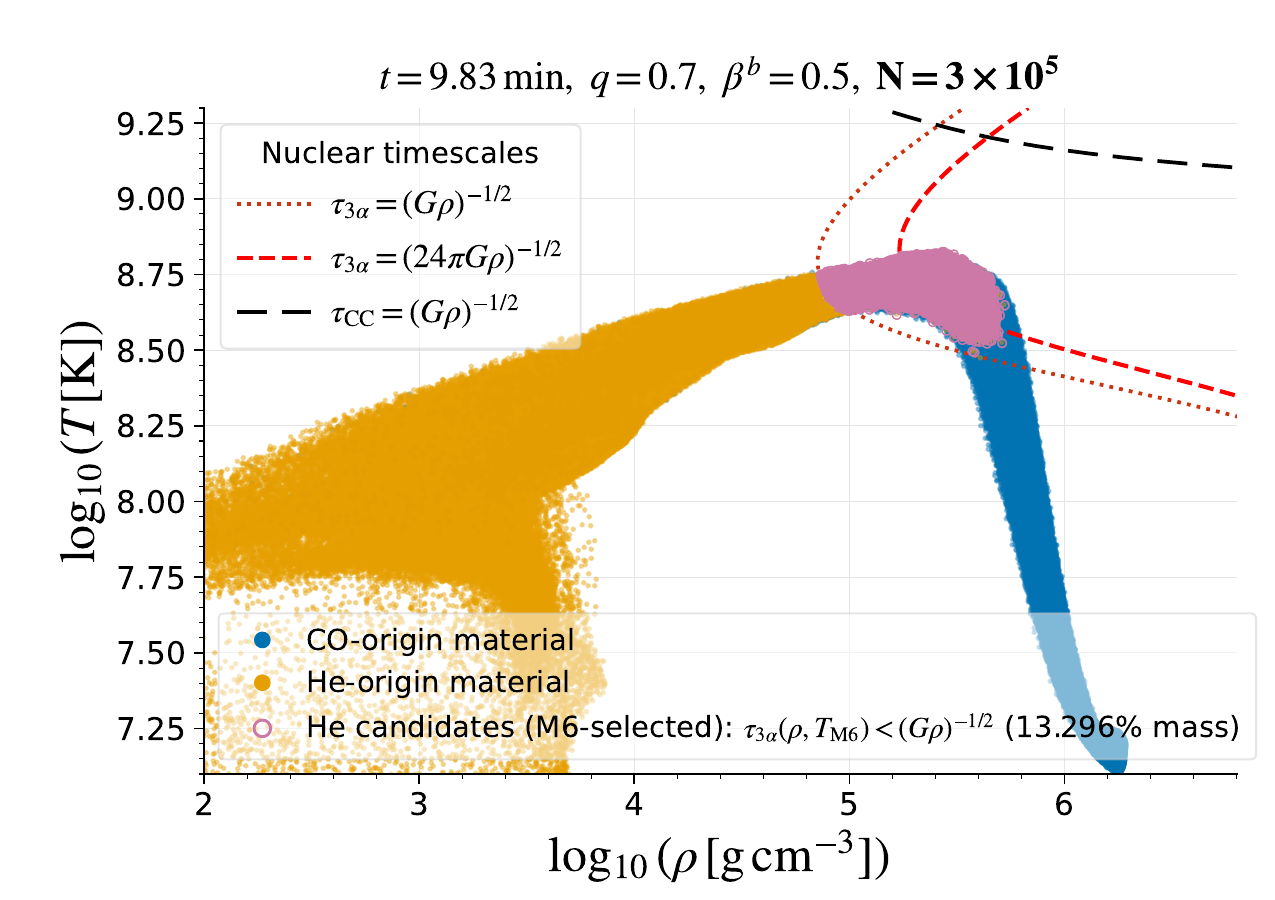}{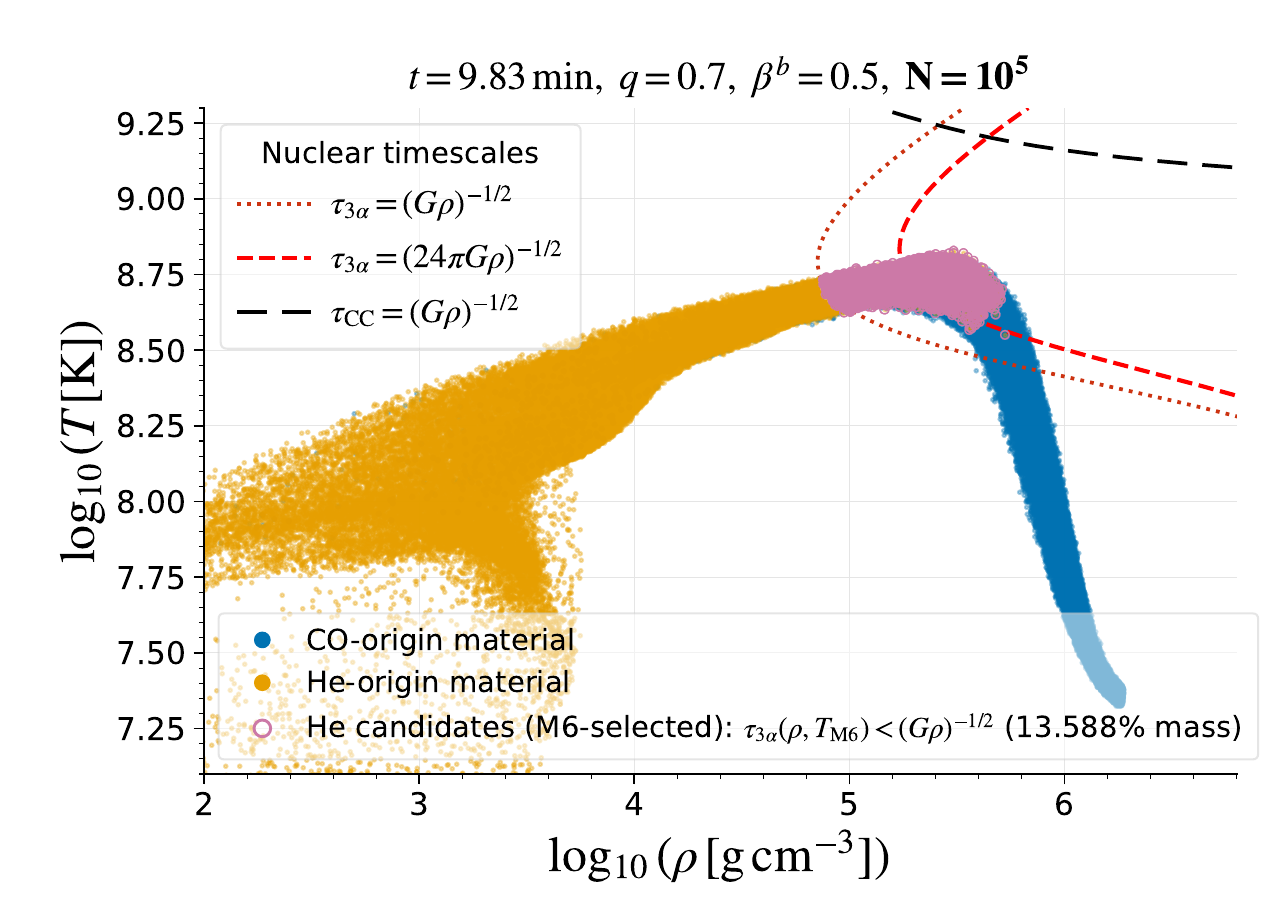}
	\caption{ Density--temperature distributions for $q=0.7$, $\beta^b=0.5$ at $t/P_0\simeq3$, at high (\textbf{left}) and fiducial
	(\textbf{right}) resolution, with symbols as in Figure~\ref{fig:rho_T}. The He-candidate fractions are $13.30\%$ and $13.59\%$.
	}
	\label{fig:convergence7}
\end{figure}

\begin{figure}[!htbp]
	\epsscale{1.17}
	\plottwo{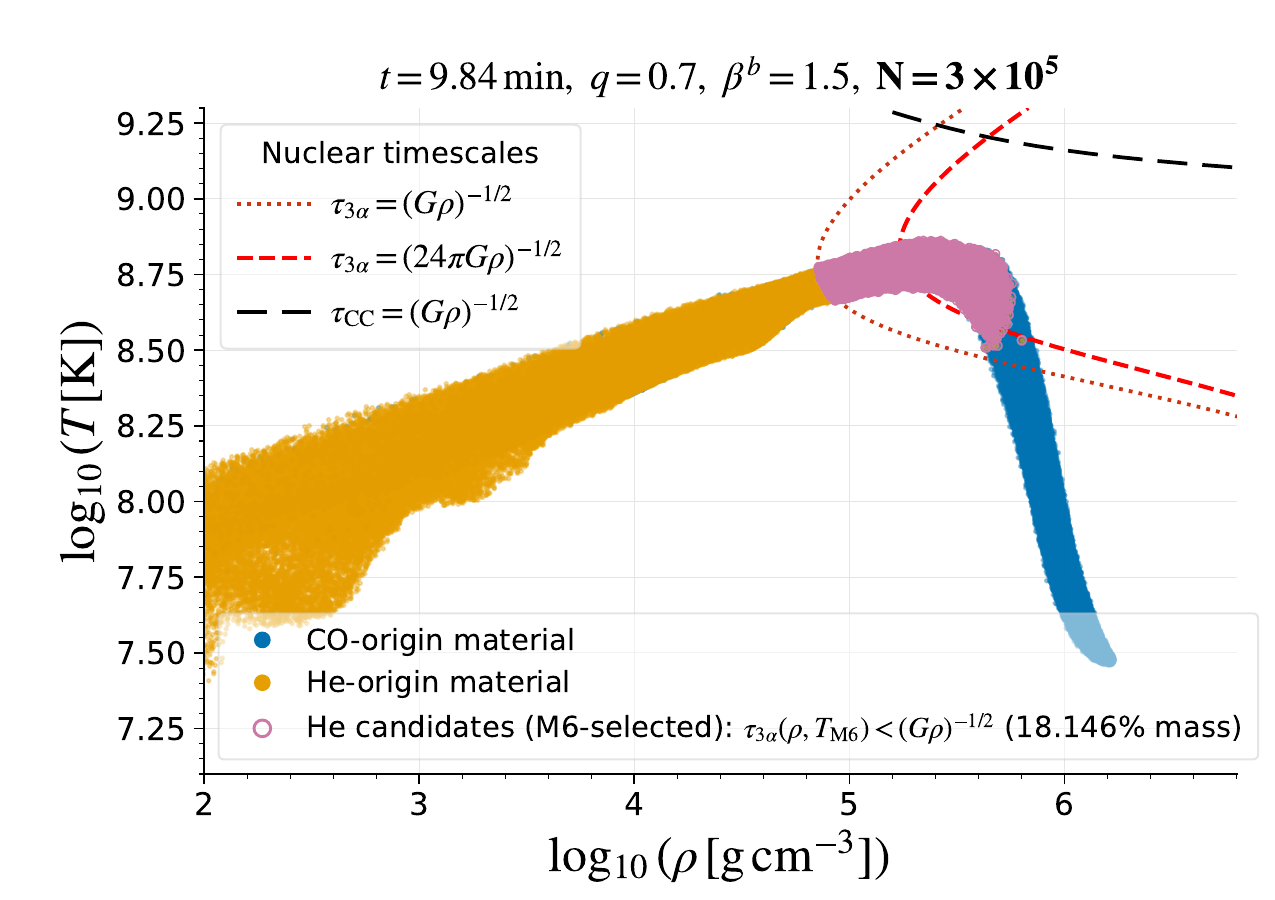}{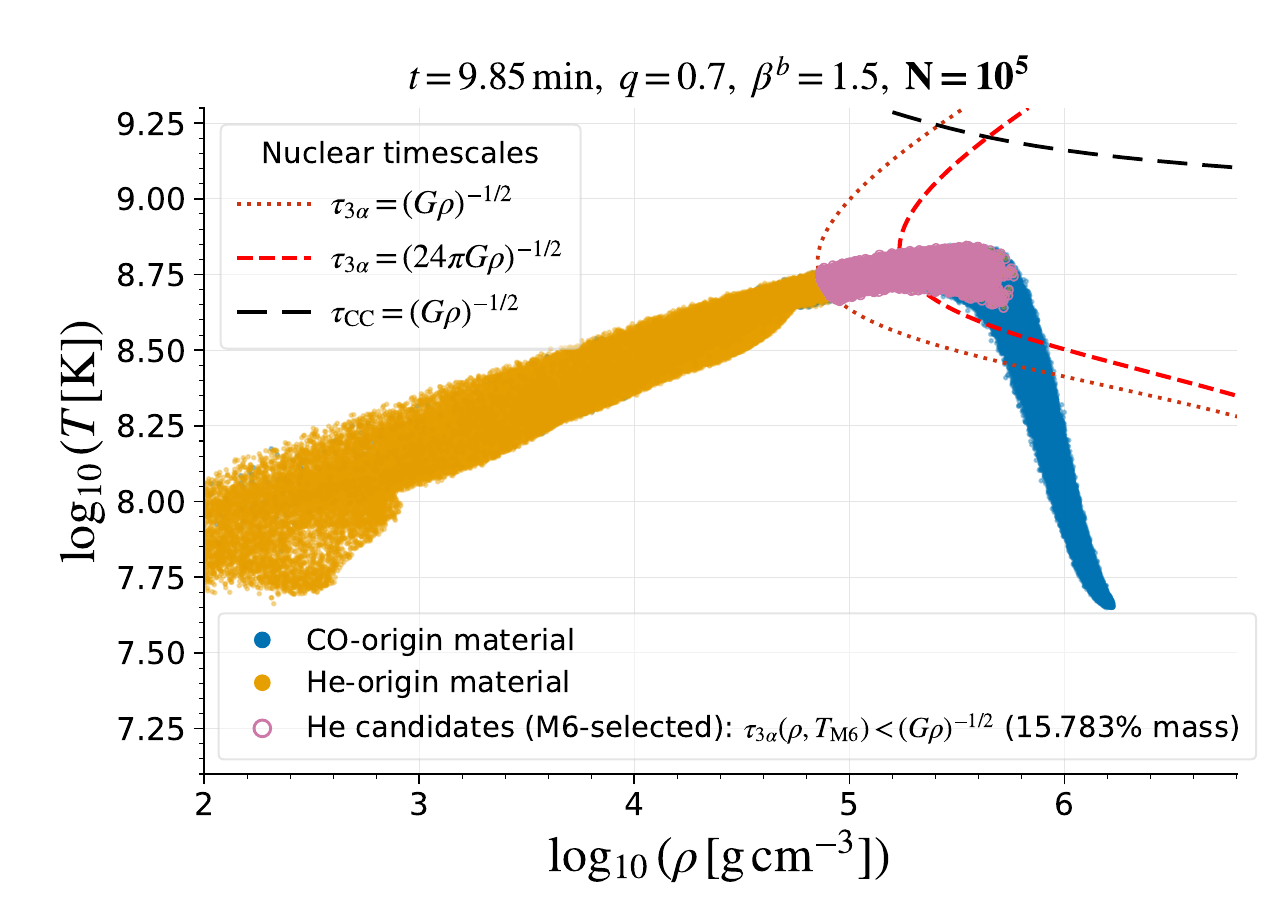}
	\caption{ As Figure~\ref{fig:convergence7}, for $\beta^b=1.5$. The He-candidate fractions are $18.15\%$ and $15.78\%$.
	}
	\label{fig:convergence8}
\end{figure}

We repeated representative calculations with $3\times10^5$ SPH
particles per WD ($6\times10^5$ in total), three times the fiducial
particle number. Particle masses are uniform within each WD but differ
between the two components. In Figures~\ref{fig:convergence1}--
\ref{fig:convergence8}, darker and lighter curves denote the high- and
fiducial-resolution calculations, respectively; solid and dashed curves
denote $\beta^b=0.5$ and $1.5$ where applicable. Time is normalized by
$P_0$ unless stated otherwise. The two resolutions reproduce the same
overall orbital, mass-transfer, and angular-momentum evolution, with
modest quantitative differences developing primarily during the
strongly nonlinear late-time phase.

\end{document}